\documentclass[preprint,prd,showpacs]{revtex4-1}
\usepackage{graphicx}

\newcommand{\pt}{\ensuremath{p_T}}
\newcommand{\bbbar}{\ensuremath{b\bar{b}}}

\newcommand{\ttbar}{\ensuremath{t\bar{t}}}
\newcommand{\tth}{\ensuremath{\ttbar h}}
\newcommand{\met}{\ensuremath{E_{T}^{miss}}}
\newcommand{\tev}{~\mathrm{TeV}}
\newcommand{\ifb}{~\mathrm{fb}^{-1}}
\newcommand{\kt}{k_{T}}
\newcommand{\et}{E_{T}}

\newcommand{\figurewidth}{0.48\textwidth}
\newcommand{\trimmedgraphic}[2][]{
  \includegraphics[width=\figurewidth,trim={0.5cm 1cm 1.5cm 1cm},clip,#1]{#2}
}
\newcommand{\trimmedgraphicsmall}[2][]{
  \includegraphics[width=2in,trim={0.5cm 1cm 1.5cm 1cm},clip,#1]{#2}
}

\newcommand{\trimmedgraphicverysmall}[2][]{
  \includegraphics[width=1.5in,trim={0.5cm 1cm 1.5cm 1cm},clip,#1]{#2}
}

\newcommand{\trimmedstack}[2][]{
  \includegraphics[width=\figurewidth,trim={0.5cm 1cm 1.1cm 1cm},clip,#1]{#2}
}

\begin{document}
\title{Machine learning techniques in searches for \tth\ in the $h \rightarrow \bbbar$ decay channel}
\author{Roberto Santos}
\affiliation{Northern Illinois University}
\author{Jordan Webster}
\affiliation{Argonne National Laboratory}
\author{Soo Ryu}
\affiliation{Argonne National Laboratory}
\author{Marcus Nguyen}
\affiliation{Northern Illinois University}
\author{Jahred Adelman}
\affiliation{Northern Illinois University}
\author{Sergei Chekanov}
\affiliation{Argonne National Laboratory}
\author{Jie Zhou}
\affiliation{Northern Illinois University}

\date{\today}
\begin{abstract}
Study of the production of pairs of top quarks in association with a Higgs boson is one of the primary goals of the Large Hadron Collider over the next decade, as measurements of this process may help us to understand whether the uniquely large mass of the top quark plays a special role in electroweak symmetry breaking. Higgs bosons decay predominantly to \bbbar, yielding signatures for the signal that are similar to \ttbar\ + jets with heavy flavor. Though particularly challenging to study due to the similar kinematics between signal and background events, such final states (\ttbar\bbbar) are an important channel for studying the top quark Yukawa coupling. This paper presents a systematic study of machine learning (ML) methods for detecting \tth\ in the $h \rightarrow \bbbar$ decay channel. Among the eight ML methods tested, we show that two models, extreme gradient boosted trees and neural network models, outperform alternative methods.  We further study the effectiveness of ML algorithms by investigating the impact of feature set and data size, as well as the structure of the models. While extended feature set and larger training sets expectedly lead to improvement of performance, shallow models deliver comparable or better performance than their deeper counterparts.  Our study suggests that ensembles of trees and neurons, not necessarily deep, work effectively for the problem of \tth\ detection.

\end{abstract}

\maketitle

\section{Introduction}
One of the key avenues of study for Higgs boson physics at the LHC is observing the interaction of the Higgs boson with top quarks; the top quark is the most massive particle ever observed in nature, with a mass over 180 times that of the proton~\cite{PhysRevLett.74.2626,PhysRevLett.74.2632,Gerber:2014xea}. This large mass of the top quark, and how it is derived from the Higgs sector, is an important area of exploration for the LHC. Unfortunately, roughly only one out of every 100 Higgs bosons at the LHC is produced in association with a pair of top quarks ($\tth$, denoted in this paper as signal)~\cite{Dittmaier:2011ti}, and the dominant top quark + anti-top quark ($\ttbar$, denoted as background) production is over 3 orders of magnitude more common~\cite{Czakon:2013goa}. The similar kinematics between signal and background events motivate the use of multivariate techniques to separate out these processes. 

The \tth\ process has not been definitively observed yet at the LHC, though multivariate techniques have already been used in multiple searches. Such techniques include artificial neural networks (ANN) with simple network structures~\cite{Aad:2015gra,Chatrchyan:2013yea,ATLAS-CONF-2014-011} and Boosted Decision Trees~\cite{Khachatryan:2014qaa,ATLAS-CONF-2014-011,ATLAS-CONF-2016-080}, and build on significant recent interest in particle physics in studying and adapting machine learning (ML) techniques to improve object reconstruction and identification~\cite{Roe:2004na,Aurisano:2016jvx,Aad:2014yva,Freeman:2012uf,Khachatryan:2015iwa,Aad:2015ydr,Guest:2016iqz}, triggering on interesting physics~\cite{Likhomanenko:2015aba,Neuhaus:2014yma}, and to classify events as in \tth\ analyses and other searches~\cite{Baldi:2014kfa,Baldi:2016fzo,Baldi:2014pta,Aad:2013dza,Chatrchyan:2012yca,Behr:2015oqq}. It should be noted that some \tth\ analyses also make use of ``matrix element methods,'' which attempt to estimate leading order probabilities for events to come from signal or background processes by integrating over their production probabilities in a large multi-dimensional space~\cite{Aad:2015gra,CMS:2014jga,Khachatryan:2015ila}, and are complimentary to the approaches studied here.

\begin{figure*}[!htb]
\begin{center}
\includegraphics[width=0.4\textwidth]{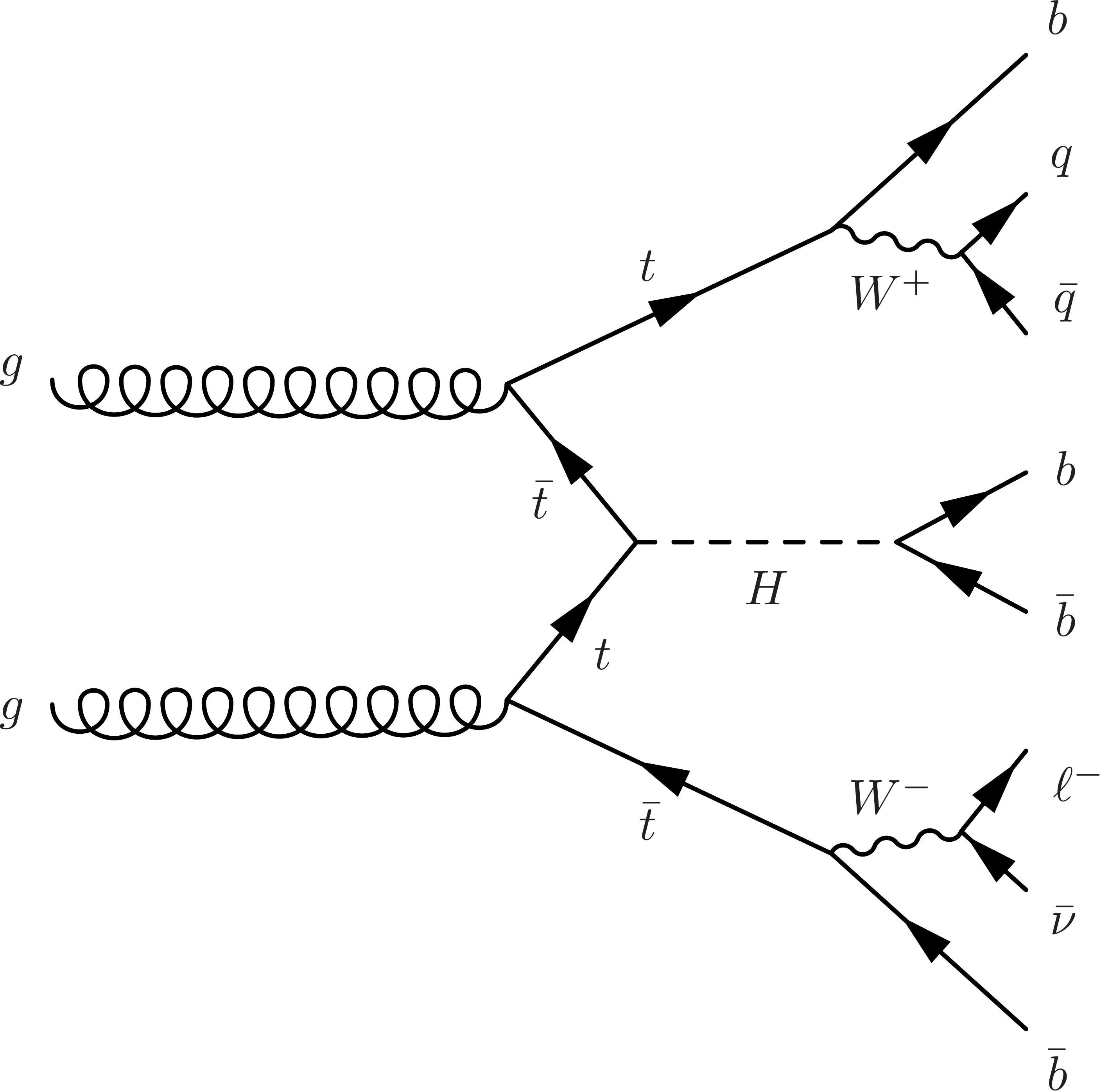}
\hspace{1cm}
\includegraphics[width=0.4\textwidth]{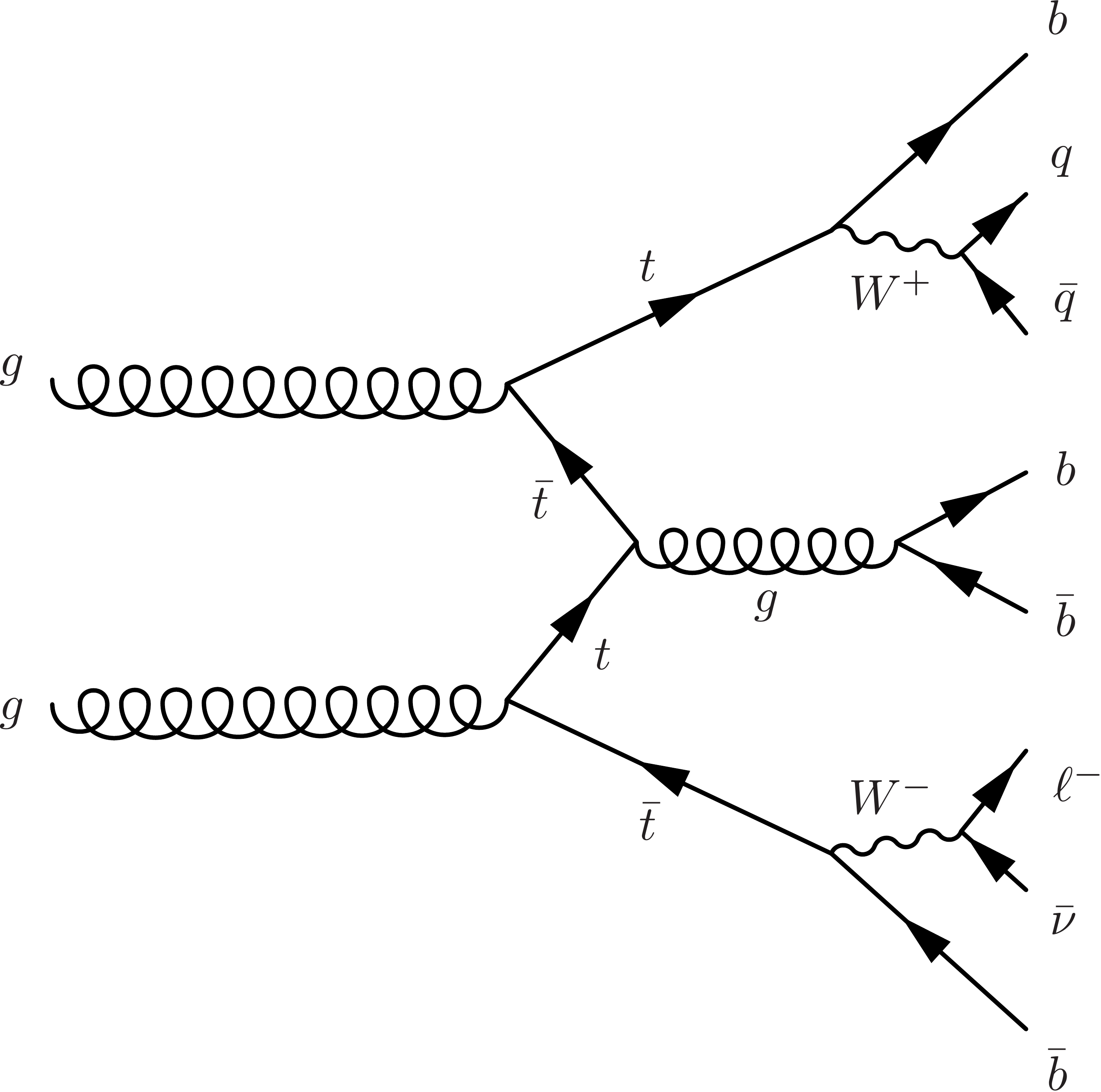}
\caption{Example Feynman diagrams for the \tth\ process of interest (left) and for the dominant \ttbar\ background produced in association with $b$ quarks (right).}
\label{fig:feynman}
\end{center}
\end{figure*}

We consider only Higgs bosons that decay via the dominant decay mode ($h \rightarrow \bbbar$). We make event selections that aim to find \ttbar\ pairs that decay semi-leptonically, though all \ttbar\ decays are included in the simulation. The $pp \rightarrow \tth\ W^+ b W^- \overline{b} \bbbar \rightarrow \ell \nu q q' \bbbar\bbbar$ production process and decay chain is a particularly interesting final state for studying the potential use of more advanced ML techniques in particle physics. The signal has eight final state objects, two intermediate $W$ boson mass resonances, two top quark mass resonances, and a dijet resonance at the Higgs boson mass that is of particular importance in separating signal and background. The neutrino is not directly detected, but instead inferred from applying conservation of momentum to the objects in the detector. In addition, most of the jets are expected to come from the hadronization and decay of $b$ quarks. An example Feynman diagram for this process is shown on the left in Figure~\ref{fig:feynman}. It is notoriously difficult to predict the rates for the dominant backgrounds in the analysis (\ttbar\ production produced in association with extra jets, particularly but not only of heavy flavor, as shown in the right in Figure~\ref{fig:feynman}), making the use of advanced ML techniques potentially useful if improved signal-background separation can be obtained. Particle physicists typically use the word ``variable'' to identify salient information about an object in the detector or an event; computer scientists and ML experts often use the word ``feature'' to reference the same item. The two words are used interchangeably here. 

%%% 2 %%%
This paper presents the first systematic studies of ML methods trained to separate the \tth\ process from the dominant $\ttbar$ background. This is a compelling use case for ML, in part because both signal and background events have rich final states with many reconstructible features. The performance of multiple ML techniques are compared, as are results using different sets of features. A number of models of gradient boosted trees are studied, as are many neural network models, including for the first time in this physics channel several models of deep networks. The wide range of models show the importance of picking optimal network structure and training parameters. The importance of the availability of large samples in order to achieve the best performance is shown. 

\section{Data sets for machine learning}
Large numbers of simulated Monte Carlo (MC) events are required for any high-energy physics analysis, in particular for those that try to separate out very small signals from large backgrounds. The MC simulations for the signal and background samples used in this paper assume $pp$ collision events at a center-of-mass energy of $\sqrt{s}=13\tev$. Data sets for the samples listed below can be found at \url{https://atlaswww.hep.anl.gov/asc/papers/1512.08928/}.

\subsection{Signal events}
Signal events are simulated using Standard Model predictions at next-to-leading order (NLO) in QCD using {\sc Madgraph5}~\cite{Alwall:2011uj}. The NLO part of the calculations is performed using aMC@NLO~\cite{Alwall:2014hca}, while the parton shower is modeled with the {\sc Herwig6}~\cite{Bahr:2008pv} generator. All decay channels for the Higgs boson and the top quarks are enabled. NN23NLO~\cite{Ball:2013hta} is used for the parton-density function (PDF), with the value of the strong coupling constant set to $\alpha_s=0.122$. The total number of signal events is 12.5 million, corresponding to an integrated luminosity of $26,975\ifb$.

\subsection{Background events}
In order to speed up the rate at which background events are simulated in the region of interest, only the $t\bar{t}+1,2,\geq 3$ parton processes are produced; the single lepton $\ttbar$ background requires at least two additional jets to enter the signal region of six or more jets, such that the zero-parton \ttbar\ sample is expected to have negligible contribution in the signal region. To test this hypothesis, a smaller sample of $t\bar{t} + 0,1,2\geq3$ partons is produced and compared to the nominal sample, and good agreement in a variety of kinematic distributions is observed after the nominal signal selection. The background process is computed at leading order with {\sc Madgraph5}~\cite{Alwall:2011uj} using tree-level matrix elements. High-order QCD effects are included using the parton-shower approach as implemented in {\sc Pythia6}~\cite{Sjostrand:2006za}, with NN23LO1 used for the PDF. The ``maxjetflavor'' option in Madgraph5 is set to 5 to allow for production of $t\bar{t}$ events in association with jets of $b$ flavor. The MLM matching scheme~\cite{Alwall:2007fs} is used to remove overlap between events, with a minimum kt jet measure of 20. The total number of background events is 16.2 million, which corresponds to an integrated luminosity of $51.8\ifb$. 

\subsection{Simulation}
All truth-level events are included in the public Monte Carlo {\sc HepSim} repository~\cite{Chekanov:2014fga}, and further processed with the Delphes 3.3 fast simulation~\cite{deFavereau:2013fsa}. This simulation uses an ATLAS-like detector geometry as included in Delphes (though results should be generic, regardless of the LHC detector), with minor modifications to reduce the event-record size and to give photons the lowest priority in object overlap removal.

Jets in the Delphes fast simulation are reconstructed with the anti-$\kt$ algorithm~\cite{Cacciari:2008gp} with a distance parameter of 0.4 using the {\sc FastJet} package~\cite{fastjet}. Calorimeter towers are used as inputs for jet reconstruction. Jets are required to have $\et>20$~GeV and $|\eta|<3$, where $E_T = E\sin\theta$ and $\theta$ is the polar angle. For jet clustering, stable particles are selected if their mean lifetimes are larger than $3\times 10^{-11}$ seconds. Neutrinos are excluded from consideration in jet clustering.

The $b-$tagging is parameterized by the Delphes 3.3 using the expected $b-$tagging efficiency and fake rates as given by the ATLAS collaboration~\cite{ATL-PHYS-PUB-2015-022}. In the nominal Delphes simulation, the $b$-tagging efficiency is parameterized as $0.80\tanh(0.003 \pt)(30/(1+0.086 \pt)$, where \pt\ is the component of the momentum in the transverse plane. Misidentification rates for light quarks and $c-$ quarks are approximately 0.002 and 0.2, respectively, with a weak dependence on the transverse energy. For this analysis, the $b$-tagging is modified to allow for multiple working points with similar functional forms but tradeoffs in terms of efficiency versus rejection of mistagged jets. Such ``continuous $b$-tagging'' can be a powerful tool in a multivariate discriminant. Figure~\ref{fig:btagging} shows the efficiencies and mistag rates for the five working points.

\begin{figure*}[!htb]
\begin{center}
\trimmedgraphicsmall{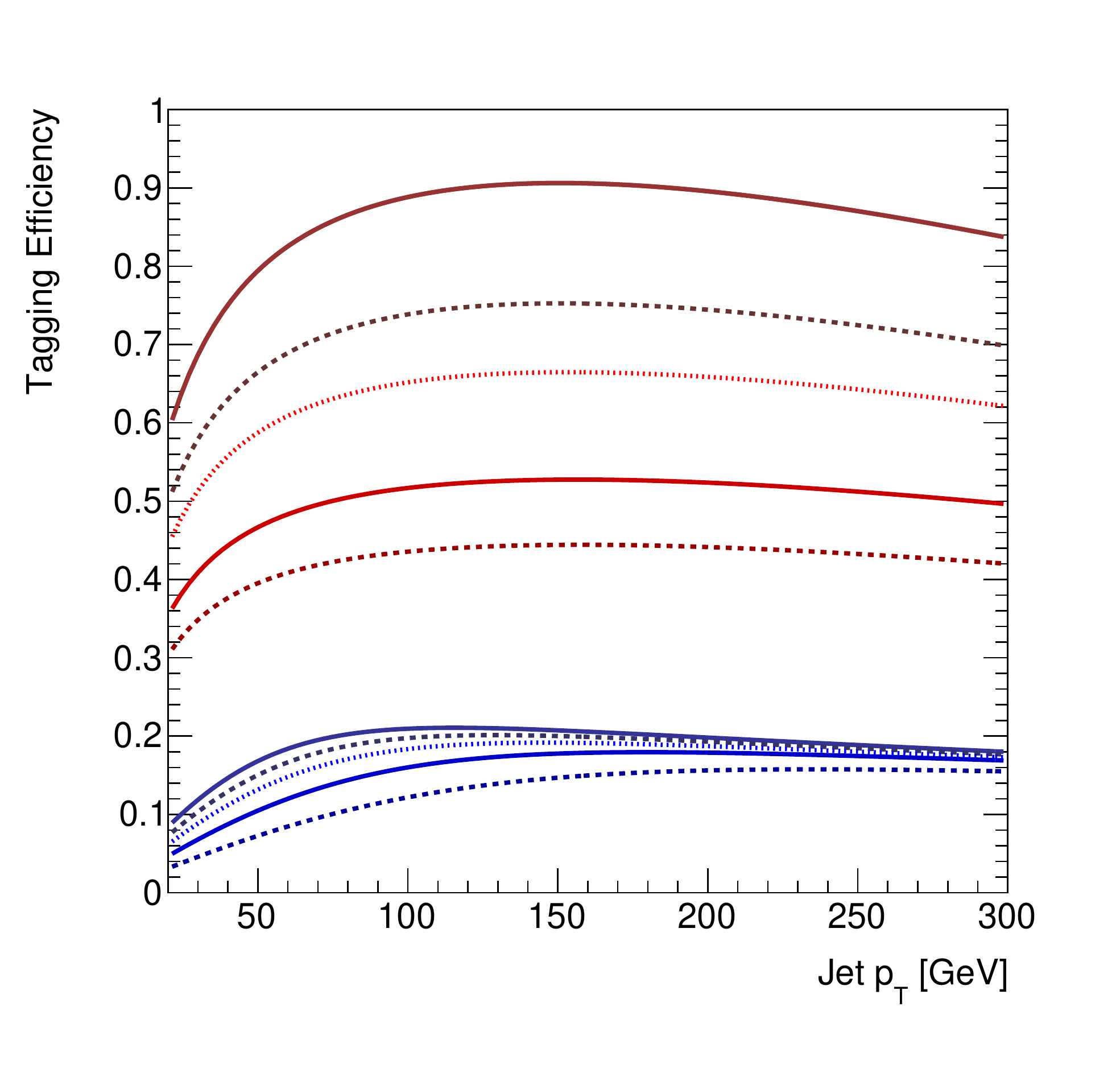}
\trimmedgraphicsmall{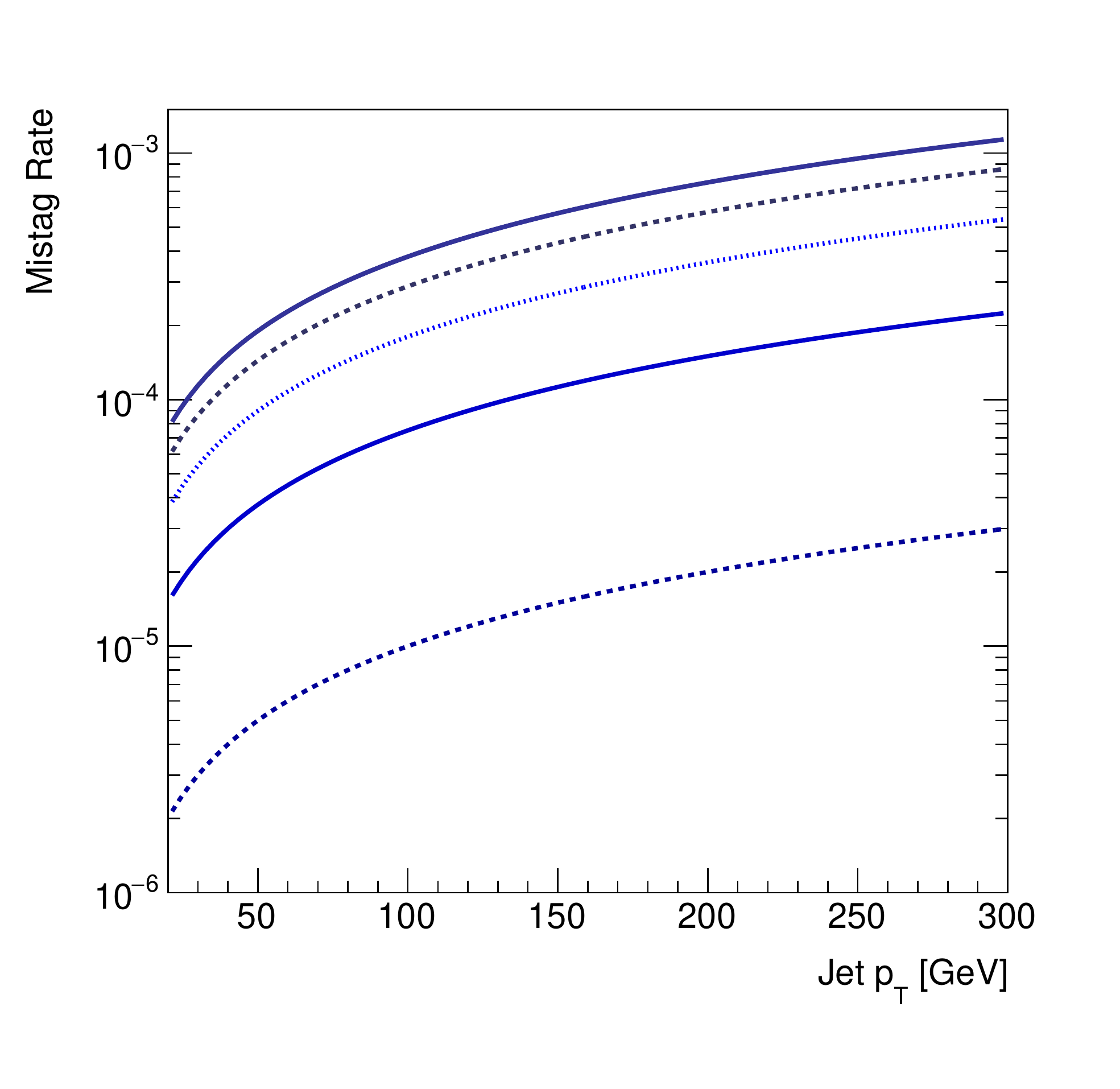}
\caption{Left: Parameterized $b$-tag (red) and $c$-tag efficiency (blue) as a function of jet \pt\ for the five working points considered in this analysis. Right: The mistag rate for these same five working points.}
\label{fig:btagging}
\end{center}
\end{figure*}

\subsection{Event selection}
\label{lab:eventSel}
Objects (jets, electrons and muons) are required to have $\pt > 20$~GeV and $|\eta| < 2.5$. The selection for both signal and background follows a loose lepton+jets selection with the requirement of extra jets and $b$-tags: exactly one electron or muon, at least six jets, and at least 3 $b$-tags using the loosest working point. The total number of signal and background events passing this selection are 1.3 million and 0.2 million, respectively.

\section{Setup for machine learning}  \label{algorithms}
To search for a process such as \tth, the main ML task is to formulate a learning model that can effectively distinguish the signal from the background.  In addition, we are also interested in understanding the full shape of the output of the ML discriminant. This is particularly important for searches for rare processes such as \tth, in which the background is not {\it a priori} well understood. By including the full shape of the output of the multivariate discriminate, physicists can control some of the uncertain parameters of the background model. In other words, it is important not only to define signal-enriched regions, but also background-enriched regions with varying degrees of signal-likeness. While multivariate techniques have been attempted in searches for \tth, a systematic understanding of the ML techniques for the problem is lacking, especially in the light of recent advancements in the ML community. Experiments are thus designed with a goal of understanding the following questions:

 \begin{enumerate}
   \item  Which learning models are effective for this problem? 
    \item Which (sets of) features are helpful for learning? Is the complete set of features more helpful than using a simpler subset?
   \item How does the size of the data set impact learning, particularly for the neural net models?  
   \item  Do deeper models, such as neural networks of more layers or deeper trees provide any advantage over the shallow ones for the given problem?
 \end{enumerate}

To answer the above questions, we investigate several ML algorithms with a focus on neural nets.  Specifically,  five ML algorithms --- Extreme Gradient Boosting (XGBoost), nearest neighbor,  decision tree,  naive Bayes and random forest  --- plus three variants of neural network models are used, which result in a total of eight models. XGBoost is a scalable implementation of gradient boosted trees \cite{chen2016xgboost}. The three variants of neural networks are described below.  These neural models have similar network structure but vary based on the optimization algorithm employed.

\begin{itemize}
\item Neural Network model 1 (NeuroBGD): This classifier is a multi-layer feed-forward neural network that uses the $\tanh$ activation function for the hidden layers, and the linear Gaussian function for the output layer.  Batch gradient descent (BGD) is used as the training rule.
  The training is stopped if the mean squared error on the validation data set failed to improve over 20 epochs.  A line search is used to determine the step size at each iteration. Due to its comparable or better performance and its stability in training, this model is the one we use to report detailed results such as the impacts of features and training size.

\item Neural Network model 2 (NeuroSGD): For this neural network classifier, the activation functions and stopping criterion are the same as model 1. Stochastic gradient descent (SGD) is used for training. 
The learning rate is initially set to 0.001, and incrementally decreases as the training progresses, down until 10$^{-6}$. The learning rate annealing is performed to give the model more sensitivity to minima as the model converges. An initial momentum of 0.9 is increased to 0.99 over the training period to speed up the training and assist in overcoming local minima, especially as the learning rate is decreased. An L2 regularization term of 10$^{-5}$ is applied to all layers, along with a drop out probability of 0.5, both of which help to avoid over-fitting.

\item Neural Network model 3 (NeuroBayes): NeuroBayes \cite{arxiv_2014_m_feindt} is a tool often used in the particle physics field, for example in recent ATLAS results~\cite{Aad:2015gra}, and is tested as an alternative model. The tool implements a feed-forward neural network using the Broyden-Fletcher-Goldfarb-Shanno method for optimization, and a Bayesian estimator deployed at the end of the network for post-processing. The specific model here consists of three layers with a single hidden layer with the number of nodes set to the number of variables + 1. A bias node is included in the input layer by default. The perceptron takes the $\tanh$ activation function for all nodes. Batch gradient descent is used for training. The number of epochs for training is set to 50. The maximum number of iterations is set to 50. The learning rate calculated by NeuroBayes is multiplied by 0.1, and the maximum learning rate is set to 0.001.

\end{itemize}

Default parameters of the XGBoost Python library are used except the tree depth and column subsampling, whose effects are discussed in Section \ref{sec:res}.  The algorithms of nearest neighbor, decision tree, naive Bayes, and random forest use the Weka library \cite{Hall+FHPRW:2009} with I/O wrappers provided in BIOCAT platform \cite{BIOCAT}. Unless specified, the default parameter setup of Weka is used.  The random forest classifier consists of 10 trees. Classification tree J48 is used as the decision trees classifier.  For neural network models 1 and 2, we make use of the neural network implementation in Theano 0.8.2 and Pylearn2 \cite{2016arXiv160502688short}\cite{pylearn2_arxiv_2013}, where Pylearn2 is a Python-based learning library built on top of Theano. We use one GPU of NVIDIA Tesla K40 with 12GB RAM and CUDA library version 7.5. The computer has a dual Intel Xeon E5 3.4GHz CPU with 504 GB memory.  

\section{Inputs to algorithms}
A total of 597 variables are considered for use as inputs to the algorithms presented in this paper. The variables are divided into two broad categories: basic variables, and extended variables. Basic variables are simple quantities of interest about the objects in our detector; in physics parlance, they include the 4-vector of each object in the detector (corresponding to the momentum of the object in each of three directions, plus the object's mass), the \met\ from potential escaping neutrinos, the $b$-tagging information associated with each jet, and the event-level jet and b-tag multiplicities. The basic variables alone are not expected to offer significant separation power between signal and background, but in combination inside of a model such as a neural network, may be able to provide useful information. In total, a minimum of 67 basic variables are considered in each event. Figures~\ref{fig:ExampleBasicVariables1}-\ref{fig:ExampleBasicVariables3} show example signal and background distributions for some of the basic variables.

\begin{figure*}[!htb]
\begin{center}
\trimmedgraphicsmall{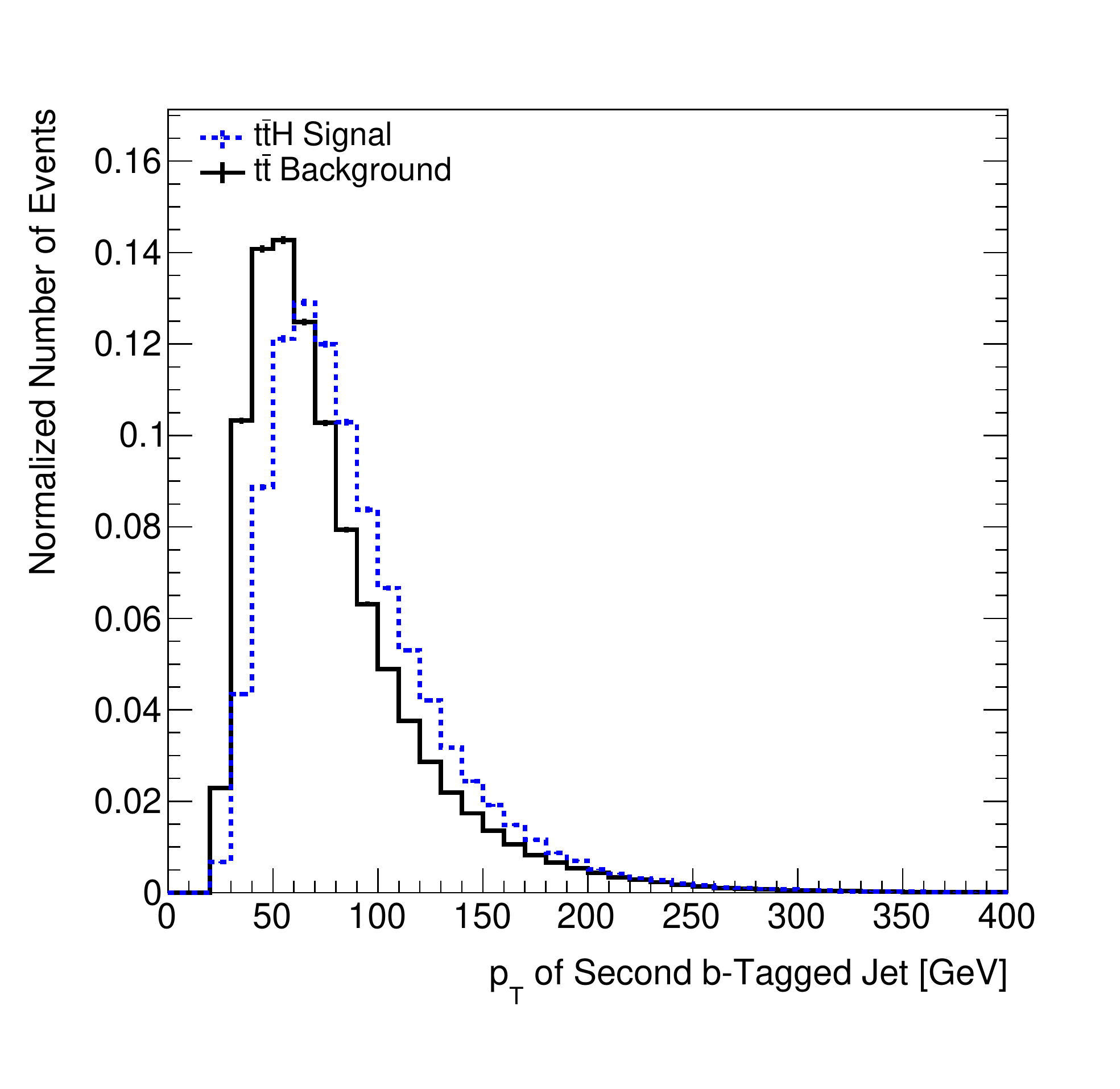}
\trimmedgraphicsmall{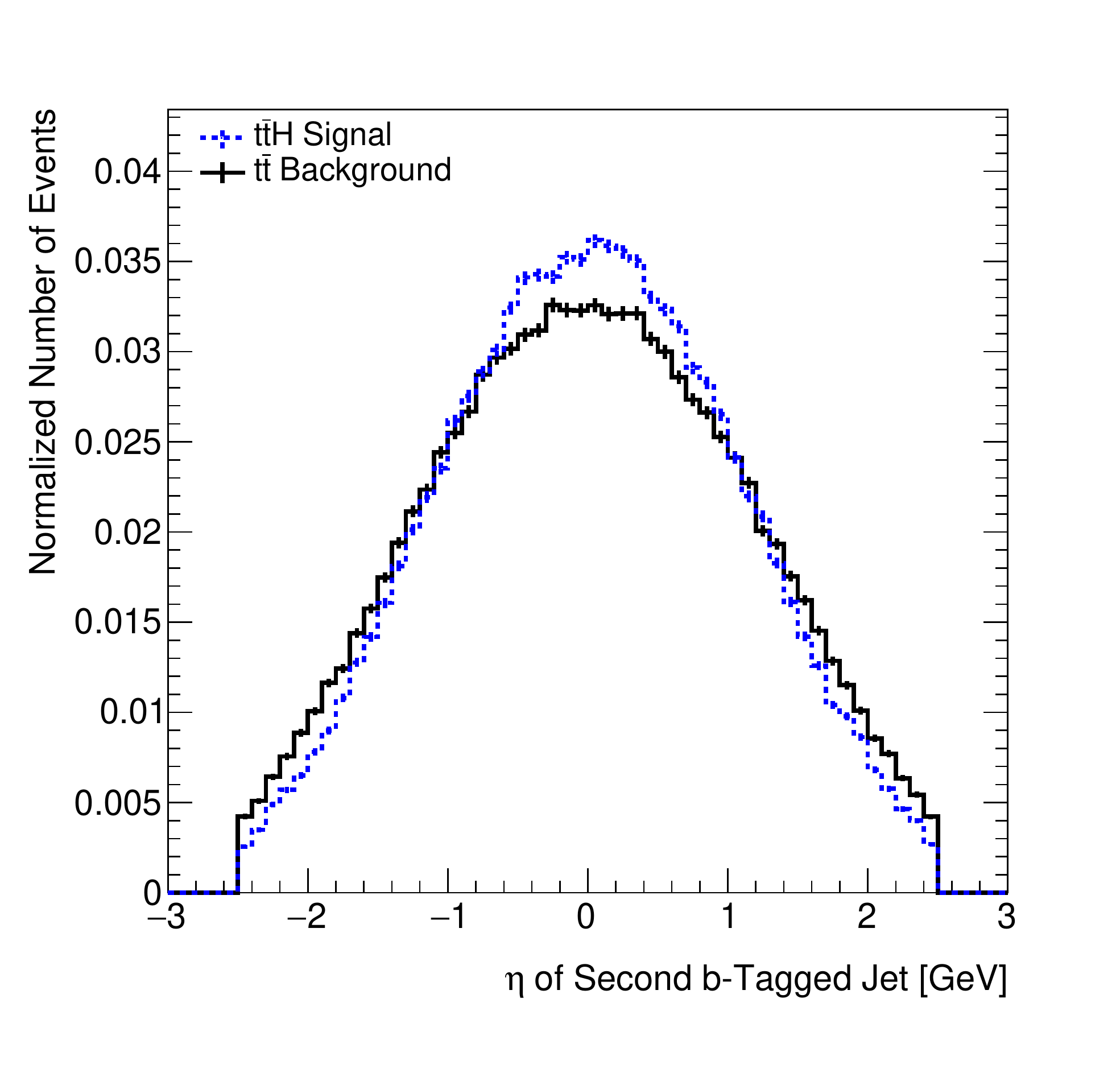}
\trimmedgraphicsmall{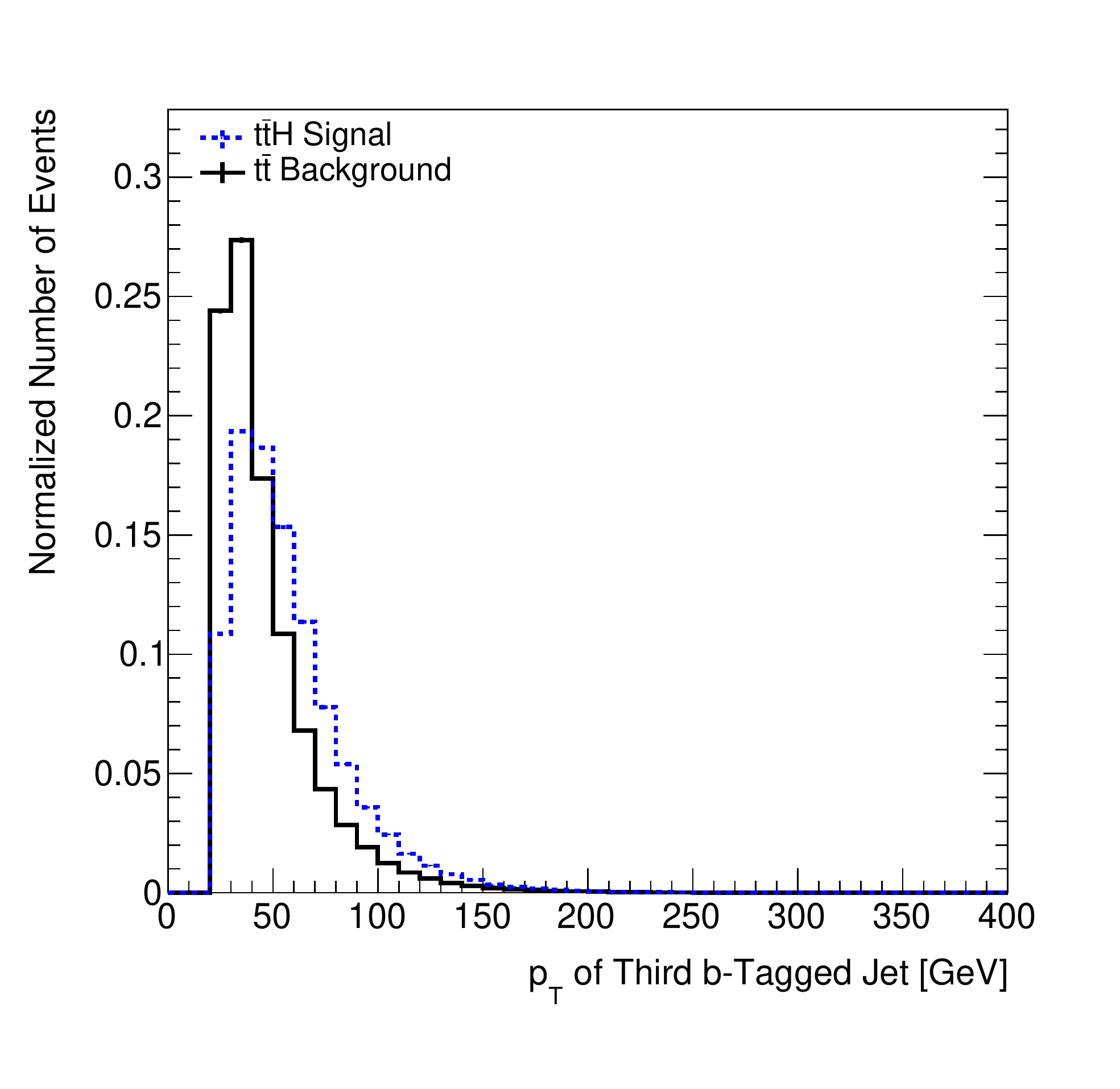}
\caption{Example signal and background distributions for three basic variables associated with b-tagged jets. Left: transverse momentum of the second hardest b-tagged jet. Middle: pseudorapidity of the second hardest b-tagged jet. Right: transverse momentum of the third hardest b-tagged jet.}
\label{fig:ExampleBasicVariables1}
\end{center}
\end{figure*}

\begin{figure*}[!htb]
\begin{center}
\trimmedgraphicsmall{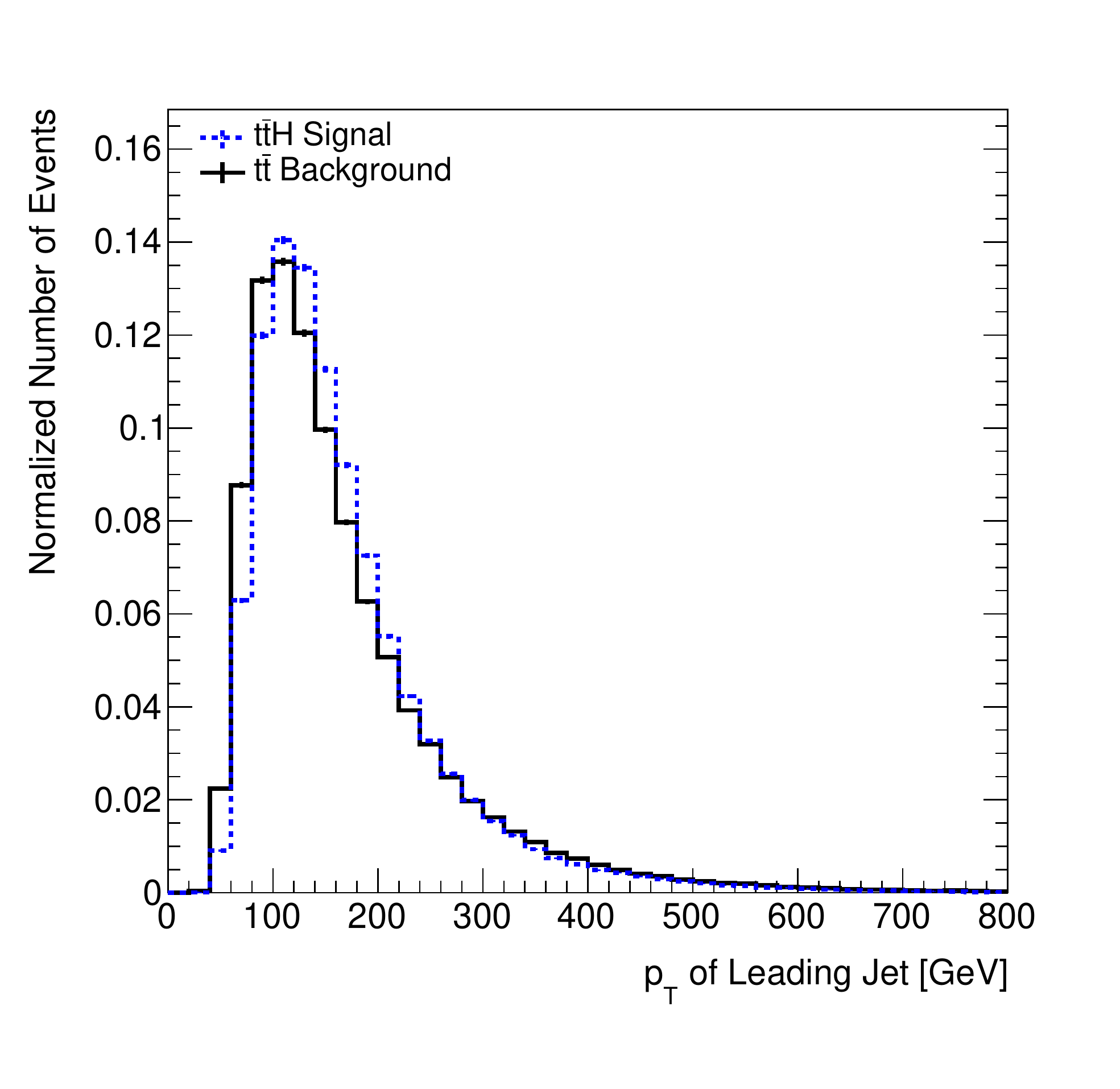}
\trimmedgraphicsmall{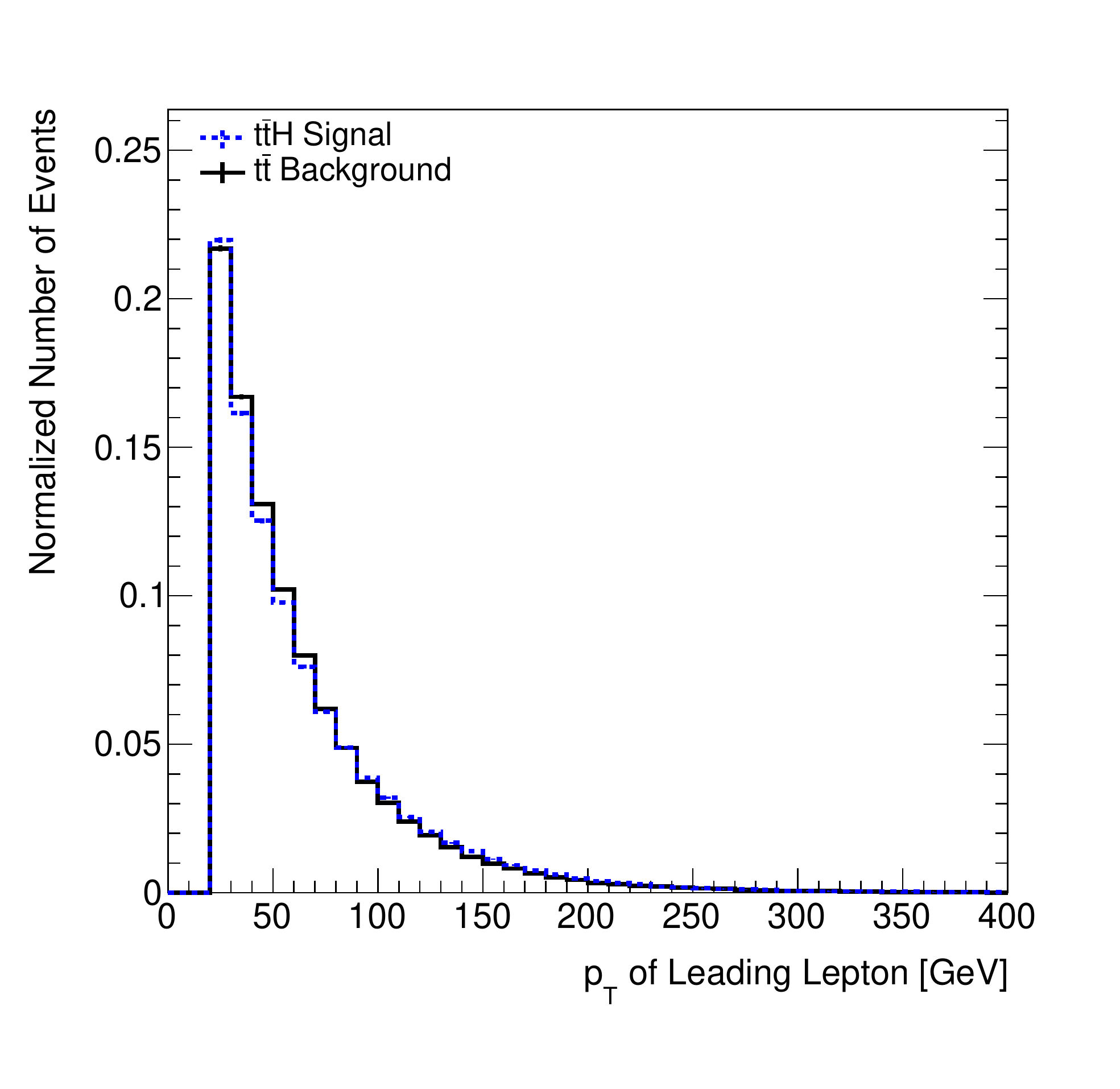}
\caption{Example signal and background distributions for two basic variables. Left: transverse momentum of the leading jet. Right: transverse momentum of the lepton.}
\label{fig:ExampleBasicVariables2}
\end{center}
\end{figure*}

\begin{figure*}[!htb]
\begin{center}
\trimmedgraphicsmall{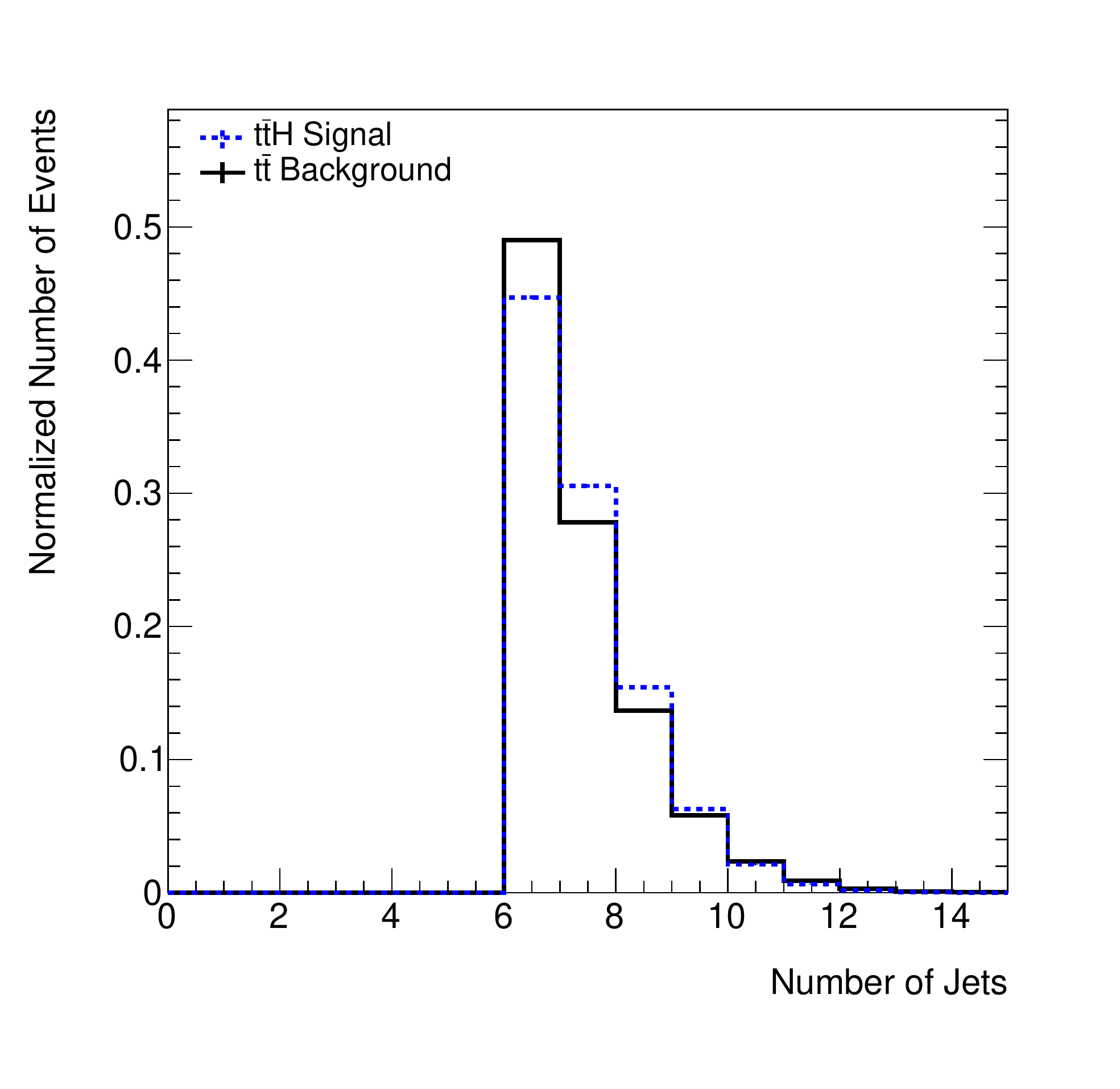}
\trimmedgraphicsmall{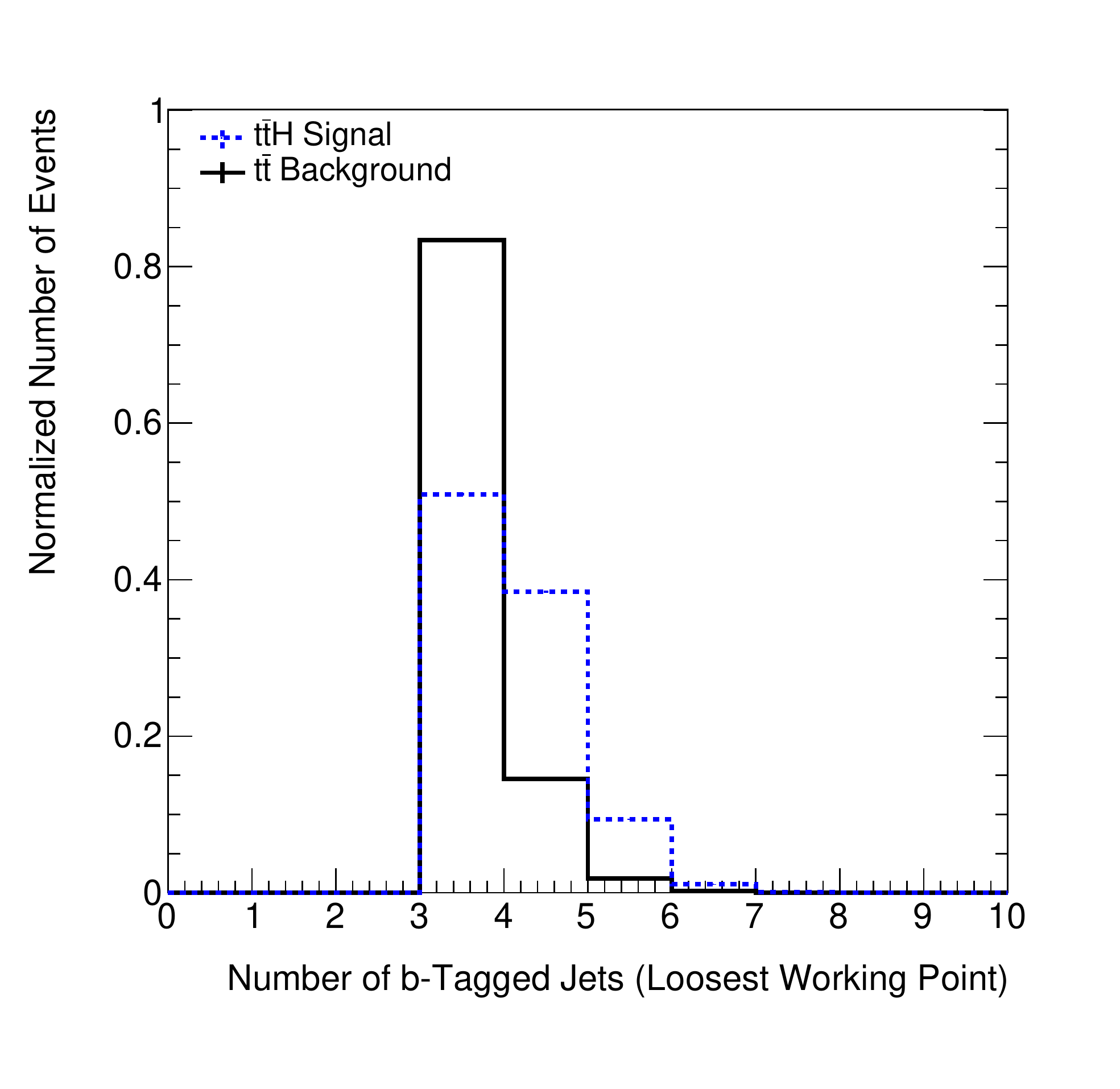}
\trimmedgraphicsmall{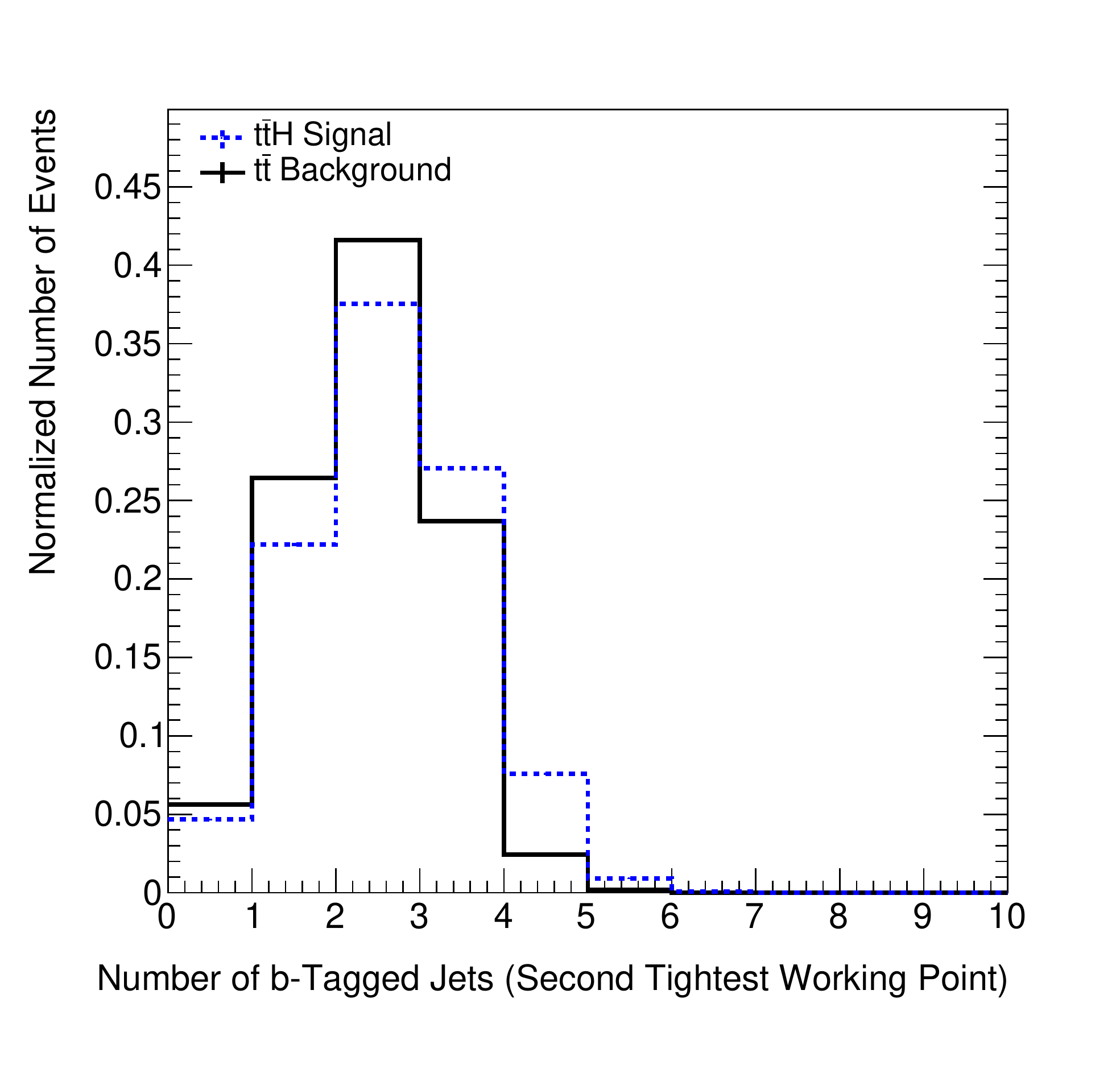}
\caption{Example signal and background distributions for three basic variables associated with jet and $b$-tag multiplicities. Left: total number of identified jets per event. Middle: total number of $b$-tags per event, as calculated using the loosest $b$-tag working point. Right: total number of $b$-tags per event, as calculated using a tighter working point.}
\label{fig:ExampleBasicVariables3}
\end{center}
\end{figure*}

In contrast to the basic features, the extended features use information about the objects in a collision inspired in some ways by physics, typically by combining multiple pieces of information. With the exception of matrix element-based analyses, ML techniques in searches for \tth\ production have so far typically used very few or zero basic features, and only a subset of extended variables. Such limitations are typically due to the complexity in working with more complicated algorithms and also the very large size of simulated samples required for an adequate training. One of the more useful quantities is the invariant mass of two or more objects; this can be thought of as the mass of a hypothetical particle decaying to the objects under consideration. The invariant mass of the pair of jets from a real, decaying Higgs boson would have a peak at the mass of the Higgs boson ($\sim 125$~GeV). Such a peak would not be present in the distributions of background events. A challenge for such features is how to select the pair of jets to use. Candidates include the pair of jets with the smallest angle (which can be a considered a distance) between them, the pair of jets with the mass closest to the Higgs boson mass, and the pair of jets with the highest momentum. Other physics-inspired variables include angles between objects in different planes, the number of objects with momentum above some minimum threshold, Fox-Wolfram moments~\cite{Bernaciak:2012nh}, and information on the shape of the collision (such as whether the flow of energy and momentum from the collision appears spherical or planar)~\cite{Aad:2012np}. The above quantities can be computed in multiple ways: for example, using all jets, jets+the lepton, all untagged jets, or only jets with a $b$-tag of some minimum purity, leading to many largely but not fully correlated variables. The $b$-tagging information itself can also be used, and is expected to be extremely powerful in separating signal from background. The selection in Section~\ref{lab:eventSel} is purposely kept loose, as the powerful $b$-tagging discriminants are also correlated with the momentum and direction of jets, so that too tight of a selection may remove salient information about the event. In total, 530 extended features are considered. Figures~\ref{fig:ExampleExtendedVariables1} and \ref{fig:ExampleExtendedVariables2} show example distributions of some of the extended variables with good separation between signal and background.

\begin{figure*}[!htb]
\begin{center}
\trimmedgraphicverysmall{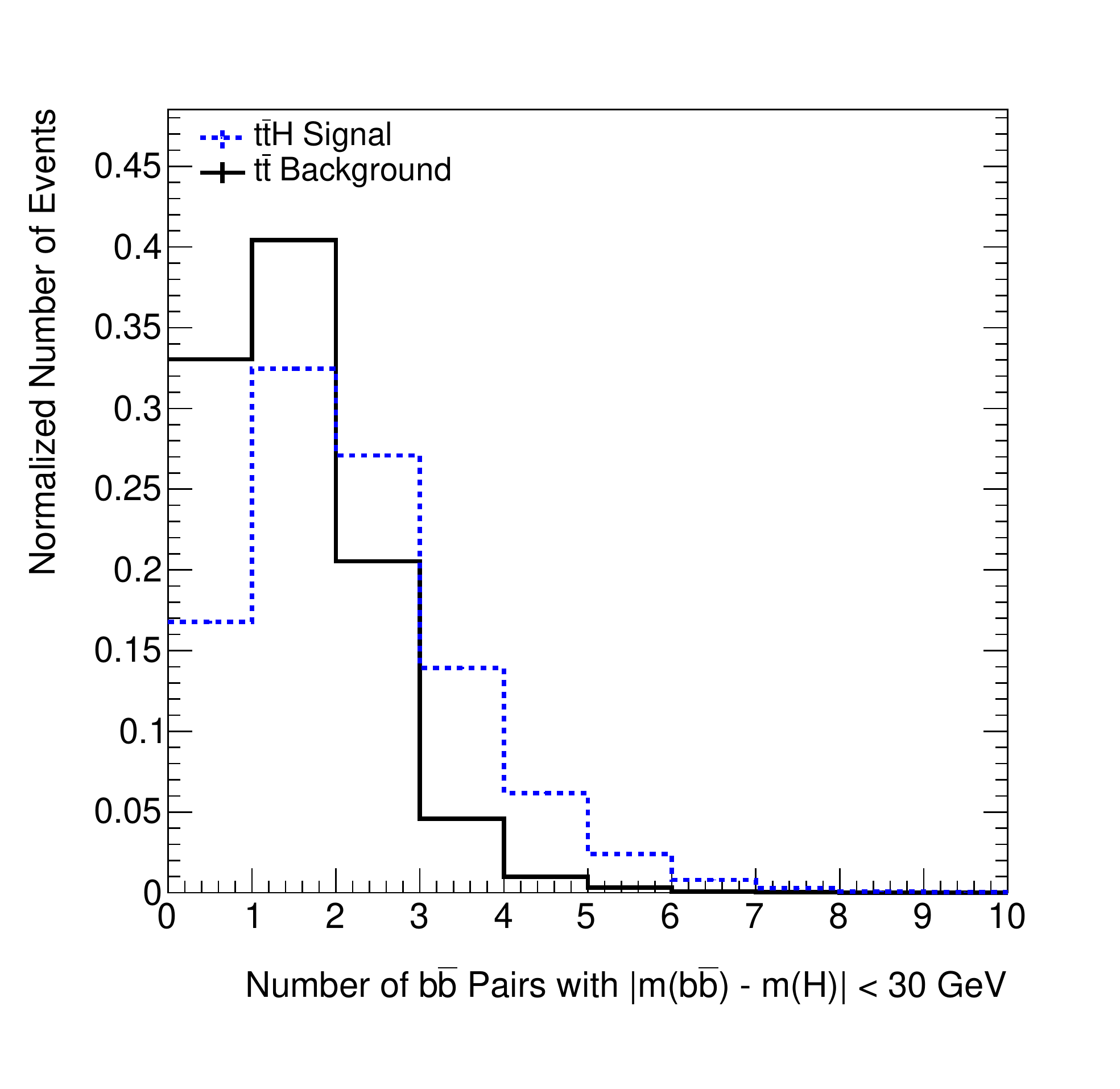}
\trimmedgraphicverysmall{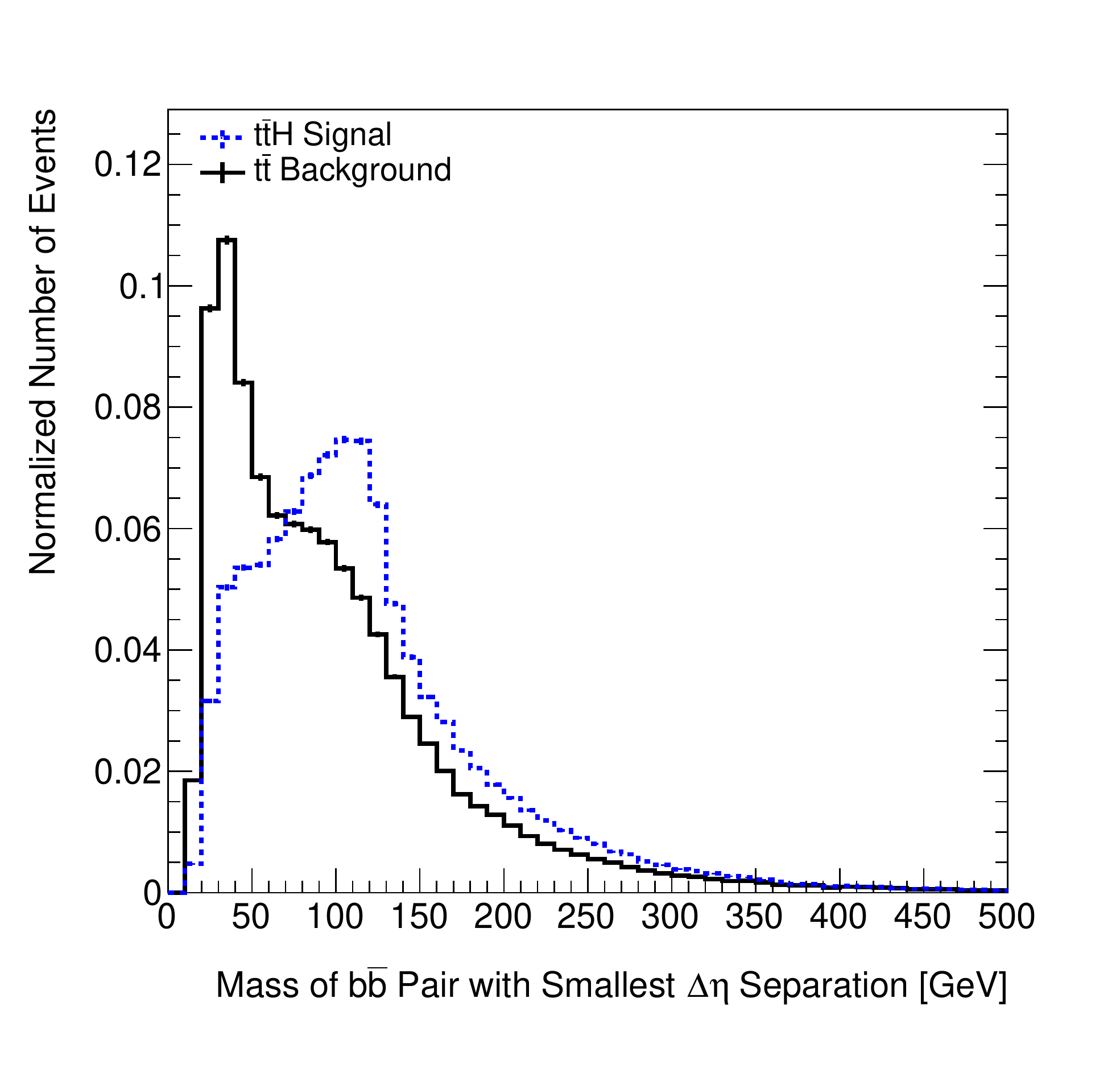}
\trimmedgraphicverysmall{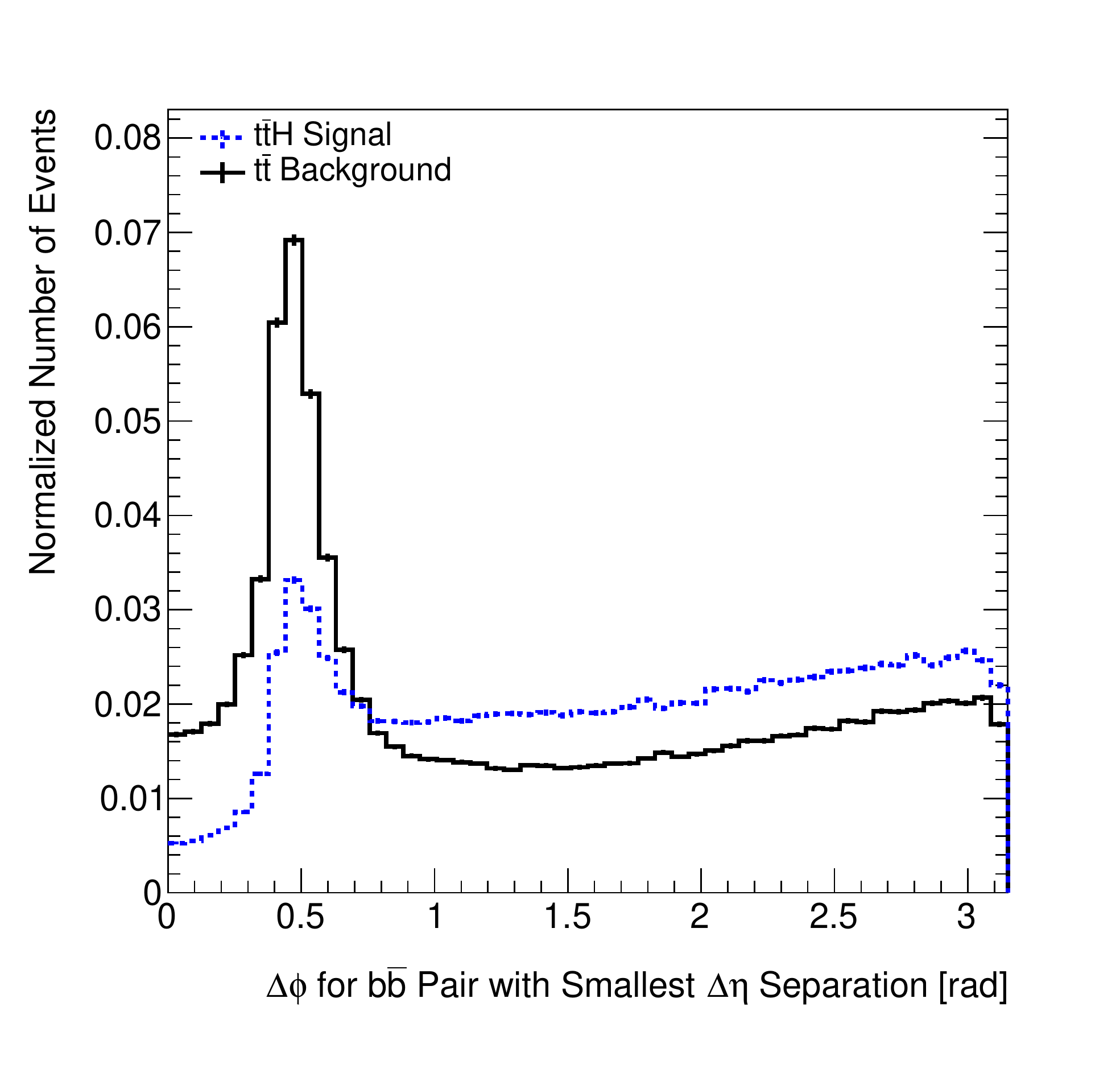}
\trimmedgraphicverysmall{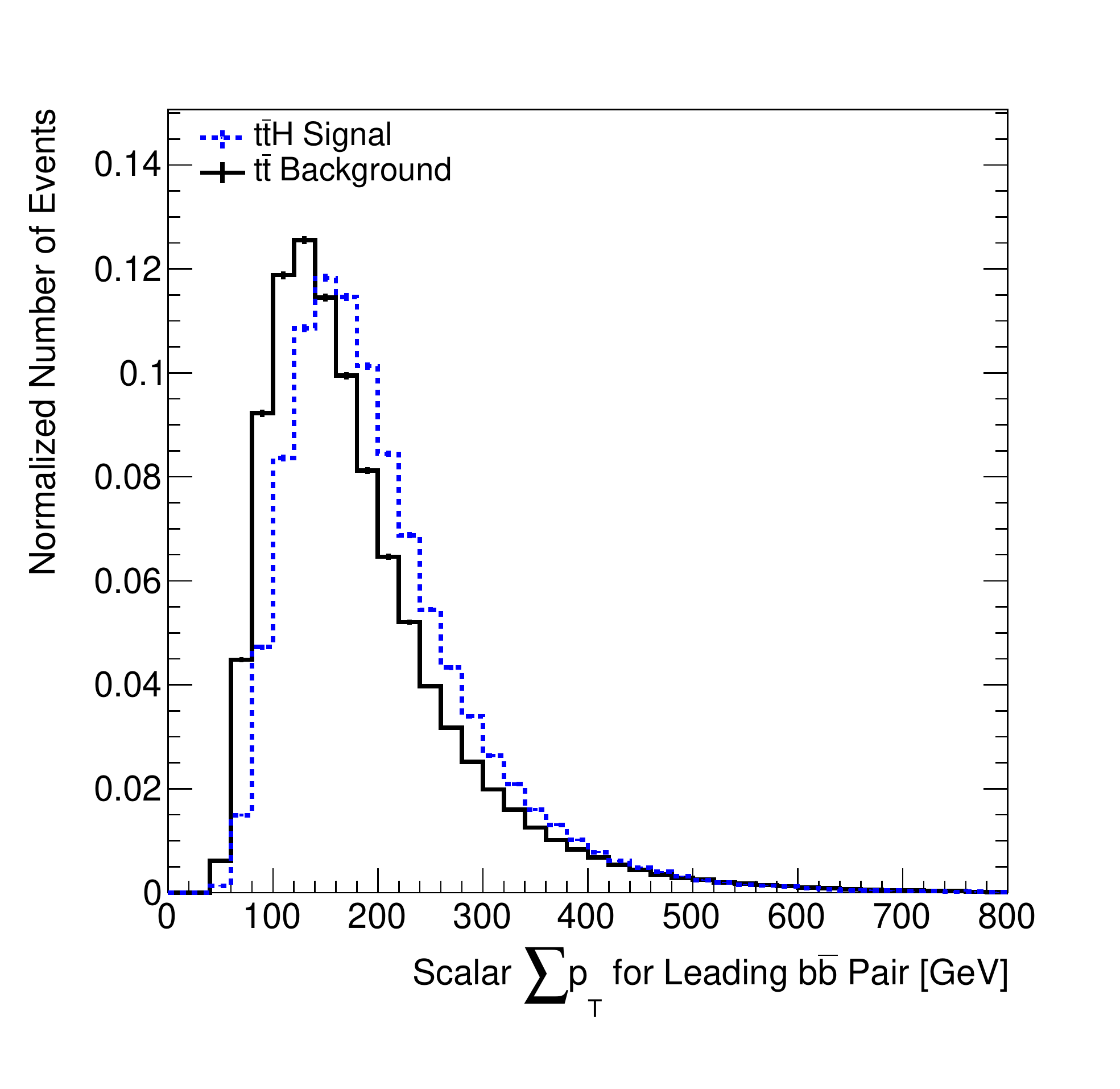}
\caption{Example signal and background distributions for four extended variables. Far left: number of $b$-tag pairs with a combined transverse mass less than 30~GeV away from the Higgs boson mass. Middle left: transverse mass of the $b$-tag pair with the smallest pseudorapidity separation. Middle right: azimuthal separation of the $b$-tag pair with the smallest pseudorapidity separation. Far right: scalar sum of the transverse momenta for the $b$-tag pair that has the highest combined transverse momentum.}
\label{fig:ExampleExtendedVariables1}
\end{center}
\end{figure*}

\begin{figure*}[!htb]
\begin{center}
\trimmedgraphicverysmall{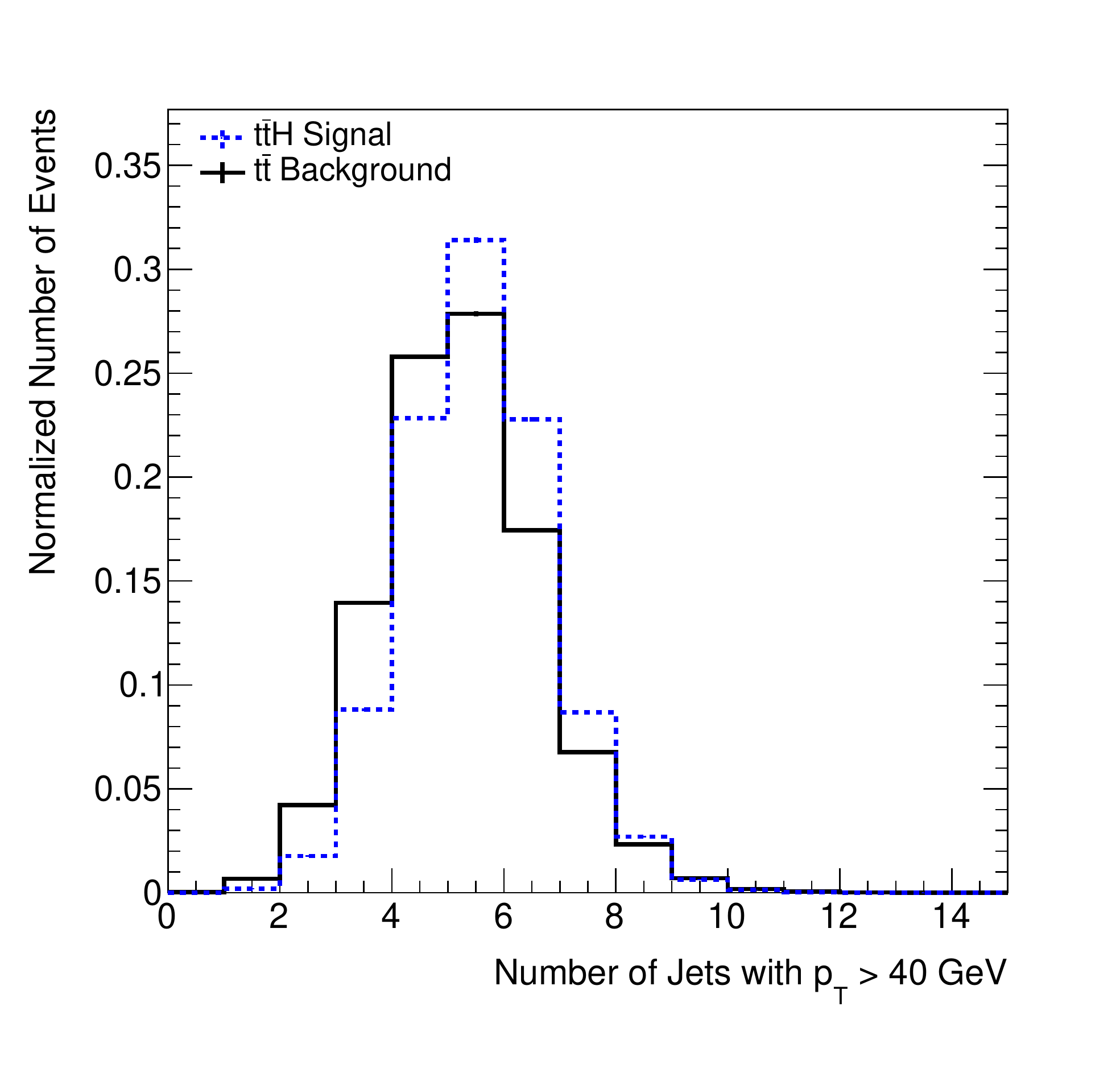}
\trimmedgraphicverysmall{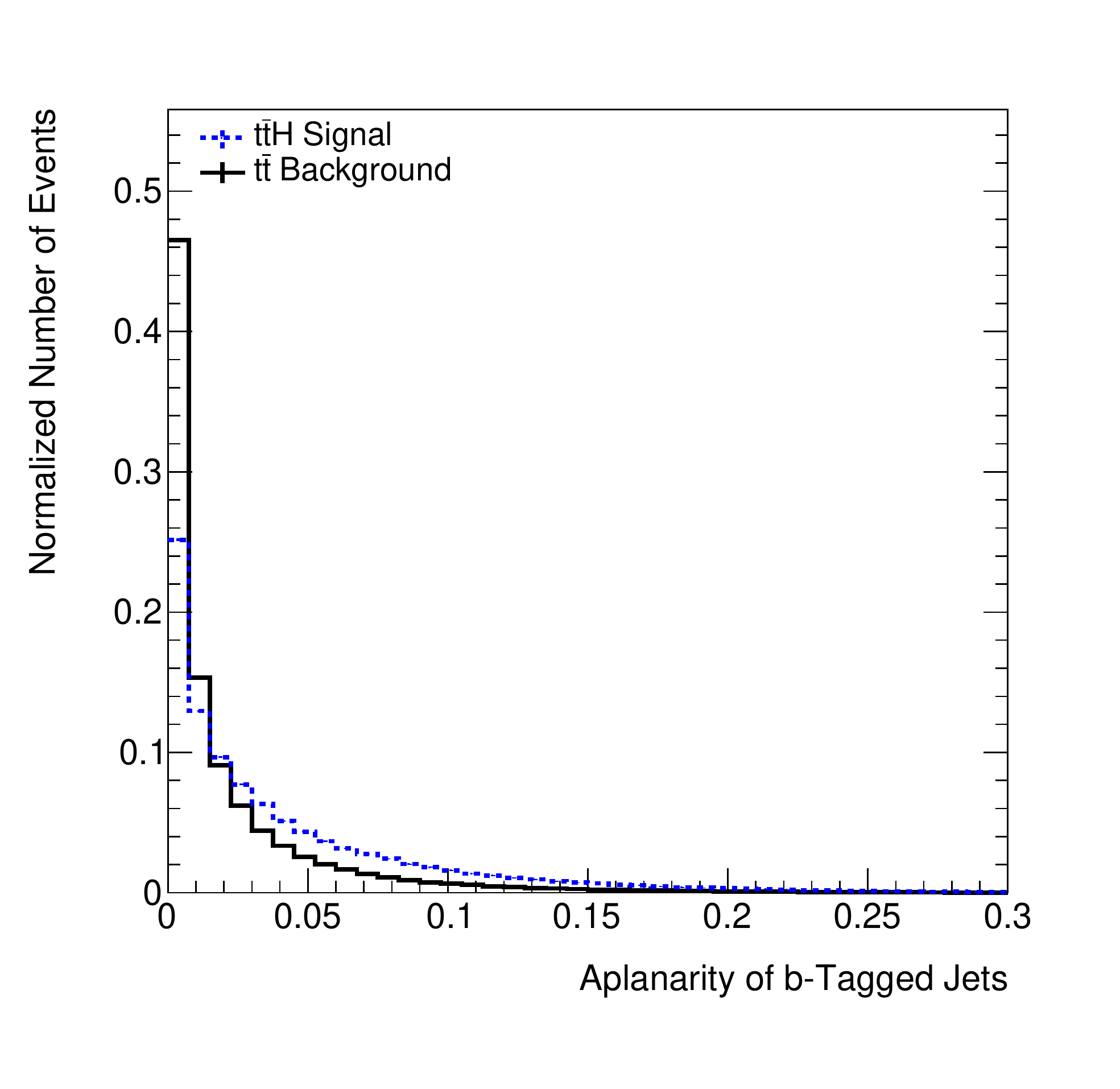}
\trimmedgraphicverysmall{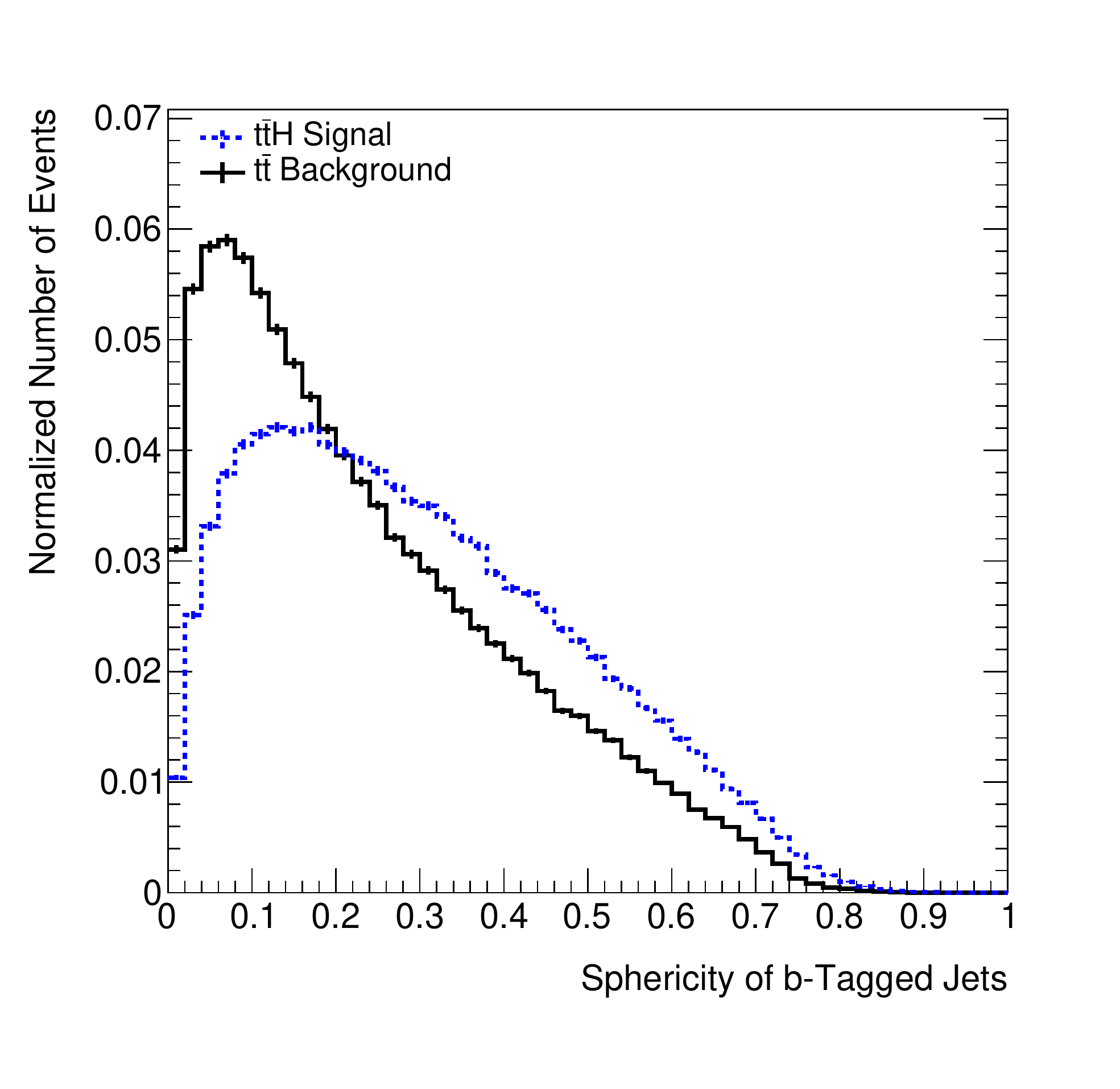}
\trimmedgraphicverysmall{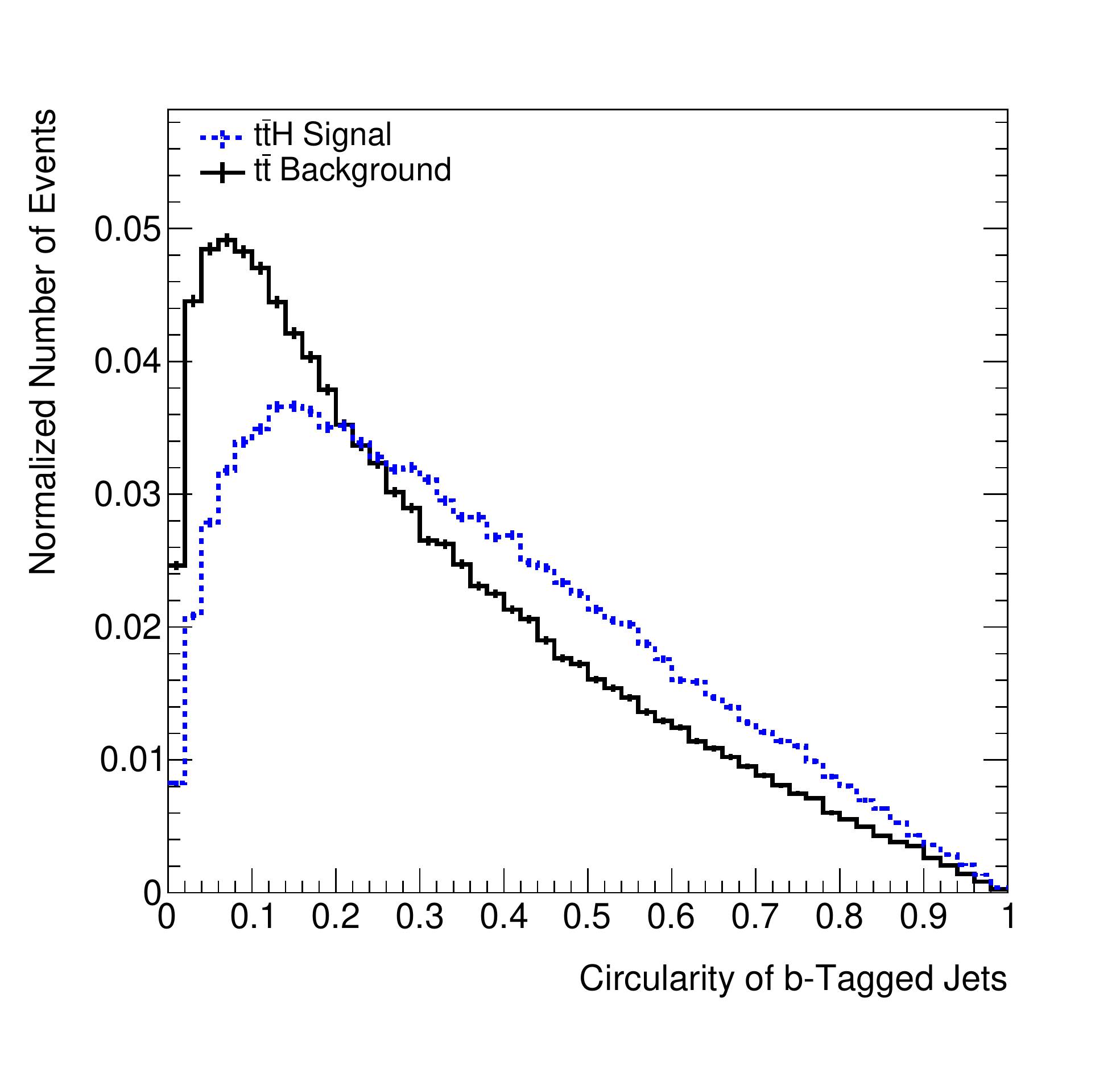}
\caption{Example signal and background distributions for four extended variables. Far left: total number of jets with transverse momentum above 40~GeV. Middle left: aplanarity of the $b$-tagged jets. Middle right: sphericity of the $b$-tagged jets. Far right: circularity of the b-tagged jets.}
\label{fig:ExampleExtendedVariables2}
\end{center}
\end{figure*}

\section{Results and discussions}  \label{sec:res}
Receiver Operating Characteristic (ROC) curves are a useful, graphical tool to indicate the performance of an algorithm, and the area under the ROC curve (AUC) is a performance measure commonly used in the ML field. By evaluating the outputs of all cut-off points,  AUC gives good insight into the effectiveness of a classifier in learning and prediction.  It is the first of two main measures used in this section and the following sections to demonstrate learning performance and compare different ML models. A second figure of merit to compare methods is the optimized value of $\frac{S}{\sqrt{B}}$, where $S$ is the expected number of signal events in a given set of data, and $B$ is the expected number of background events in that data set. This is a common figure of merit when looking for a signal on top of a background in data sets with large numbers of events, as $\sqrt{B}$ can be considered as a $1 \sigma$ uncertainty on a Poisson process with expectation value $B$. Thus, $\frac{S}{\sqrt{B}}$ is a rough estimate of the significance of a signal observation in the limit of a large number of events and no systematic uncertainties. The advantage to using this figure of merit, as opposed to other, more complicated versions for discovery, is that relative comparisons between methods are insensitive to a global scaling of the luminosity for a study. In addition, F-score, defined as the harmonic mean of the precision rate and recall rate, is also studied.

\subsection{Comparing learning algorithms}  \label{sec:algocomp}

Table~\ref{tab:compalgo} compares the effectiveness of the eight machine learning algorithms described in Section \ref{algorithms}. The results are based on a set that contains 423,173 total events, with equal contributions from signal and background. Among them, 10\% of the data set is set aside to provide an independent validation set for model training. Another 10\% is used as the testing set, and the remaining 338,541 events are used for training.
Each feature is normalized over the entire balanced data set using the z-score so that the data mean is 0 and the standard deviation is 1. An equal number of samples are randomly selected from the signal and background data pool when designating events for testing or validation.  The training set is randomly shuffled.  Experiments are conducted using the full set of 597 features, with the exception of NeuroBayes, which uses the top 200 ranked features due to the software's limit on input feature dimensionality (see section \ref{sec:features} for feature ranking).  For NeuroSGD and NeuroBGD, the network has 2 layers and 15 nodes per layer. In addition to the Tanh activation function, a Rectified Linear (ReLU) activation function is also tested for the hidden layers. ReLU takes less training time than the Tanh activation function and delivers comparable AUC.  For XGBoost, the result in Table  ~\ref{tab:compalgo} is obtained by a model of depth 1 and number of training iterations 1000, with column subsampling set to 25 percent.

\begin{table*}[!t]
\center
     \begin{tabular}{ l | l | l | l | l | l | l | l | l }
     \hline
      Algorithm & KNN  & Naive Bayes & Decision Tree & RF & NeuroBayes & NeuroSGD  & NeuroBGD & XGBoost \\ \hline\hline
        AUC &   59.9 & 71.5  & 62.3  & 78.4   &  77.7 & 78.7  & 80.0 & 80.2 \\ \hline
        F-score & 60.0 & 62.7 & 64.8 & 69.5 & 61.8 &  73.2 & 74.2 & 74.1 \\ \hline
\end{tabular}
\caption{Comparison of learning algorithms (\%). KNN: K-nearest neighbor. RF: Random Forest.}
\label{tab:compalgo}
\end{table*}

%Table~\ref{tab:compalgo} shows that the three neural network models are more effective for the given problem compared with alternatives.
Table~\ref{tab:compalgo} shows that XGBoost and the three neural network models are more effective for the given problem compared with the alternatives. Among the three variants of learning models based on neural nets,  NeuroBGD and NeuroSGD outperform NeuroBayes, a commonly used algorithm in particle physics analyses. The small advantage that NeuroBGD and NeuroSGD have over NeuroBayes remains true when identical feature sets are used between the models.
XGBoost and NeuroBGD are the top two performing models and they have nearly identical performance.
 Excluding XGBoost and the three neural learning algorithms, the random forest performs the best. A comparison among the ML methods using ROC curves is shown in Figure~\ref{fig:ROCCurveClassifiers}.

\begin{figure}[!htb]
\begin{center}
\trimmedgraphic{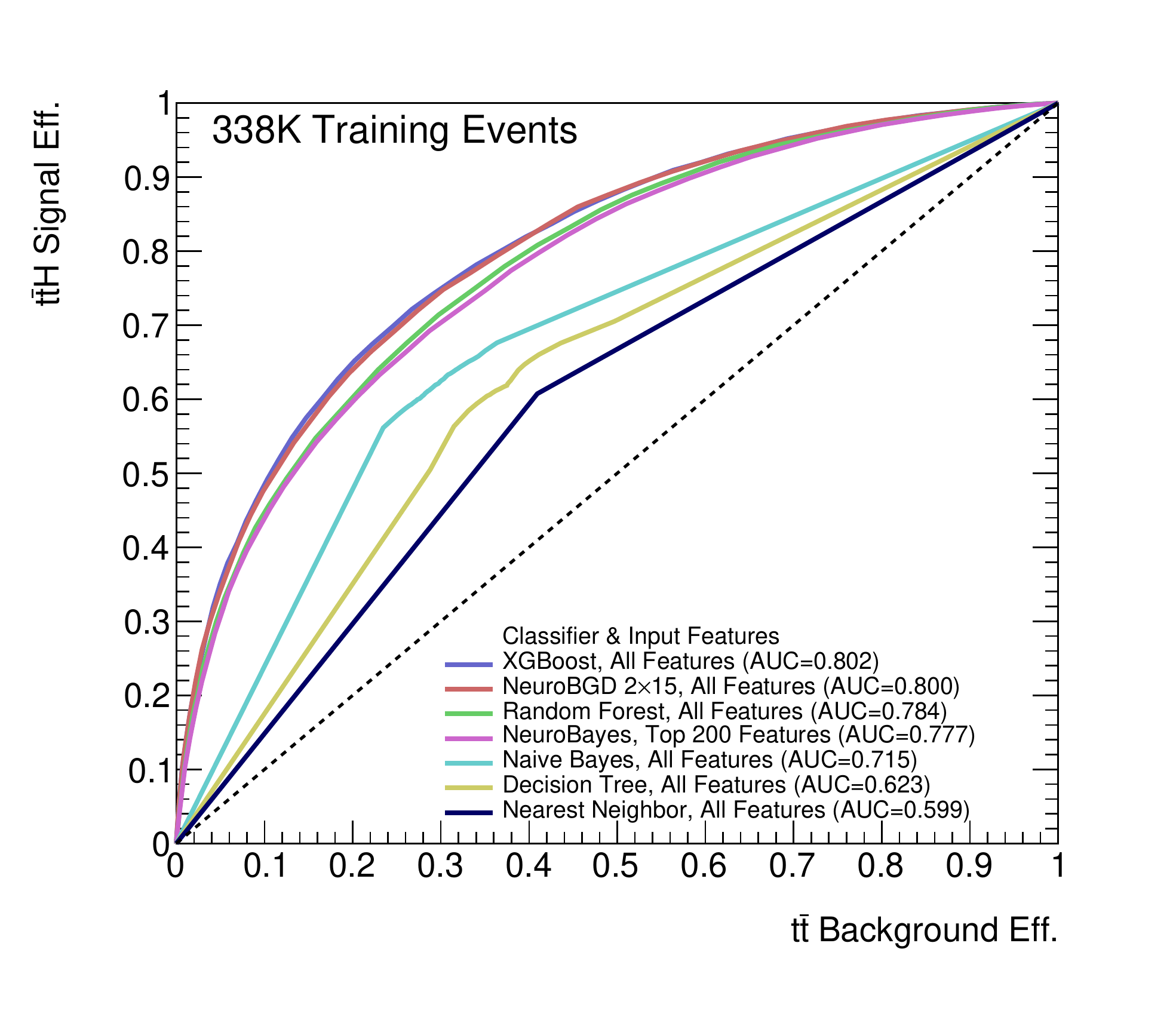}  
\caption{ROC curves from different classifiers.}
\label{fig:ROCCurveClassifiers}
\end{center}
\end{figure}

For each ML technique under study, an optimized requirement is placed on the value of the output discriminant by choosing the cut such that it gives the largest value of $\frac{S}{\sqrt{B}}$. Table~\ref{tab:compalgoSB} shows a comparison of these optimized cuts and this figure of merit. It also shows the relative increase in the amount of data that would be needed by each algorithm to achieve the same (statistical-only) sensitivity as the best performing algorithm. Since statistical sensitivity goes as the square root of the amount of data, small increases in $\frac{S}{\sqrt{B}}$ can still indicate large improvements in performance, and be quite important for experiments that can take years, if not decades, to accumulate data sets of sufficient size for discovery. For example, the best neural network structure needs 3\% more data to achieve the same performance as XGBoost, and using only the extended features requires at least nearly 18\% more data. The use of only basic features requires 51\% more data. In addition, the worse performance of algorithms such as the decision tree and random forest can be seen clearly. The optimal network (using as many inputs as possible) from NeuroBayes requires 27\% more data to achieve the same sensitivity as the leading XGBoost algorithm.

\begin{table*}[!t]
\center
     \begin{tabular}{ l | l | l | l }
     \hline
     Algorithm & Structure & Optimal $\frac{S}{\sqrt{B}}$& Relative data needed \\ \hline\hline
\hline
XGBoost (AF) & & 11.90 & -- \\
\hline
NeuroBGD (AF) & 2x15 & 11.72 &  3.0 \% \\ 
NeuroBGD (AF) & 11x300 & 11.30 &  10.8 \% \\
NeuroBGD (AF) & 2x300 & 11.25 &  11.7 \% \\
NeuroBGD (AF) & 5x300 & 11.23 &  12.3 \% \\
NeuroBGD (AF) & 8x300 & 11.21 &  12.7 \%  \\
\hline
NeuroBGD (EF) & 2x300 & 10.96 &  17.9 \% \\
NeuroBGD (EF) & 5x300 & 10.95 &  18.1 \% \\
NeuroBGD (EF) & 8x300 & 10.88 &  19.6 \% \\
NeuroBGD (EF) & 11x300 & 10.86 &  20.1 \% \\
\hline
NeuroBGD (BF) & 2x15 & 9.67 &  51.4 \% \\
NeuroBGD (BF) & 2x300 & 9.59 &  53.9 \% \\
NeuroBGD (BF) & 11x300 & 9.59 &  53.9 \% \\
NeuroBGD (BF) & 8x300 & 9.58 &  54.3 \% \\
\hline
NeuroBayes (top 200) & 201 & 10.55 &  27.0 \% \\
NeuroBayes (top 20) & 21 & 9.80 &  47.5 \% \\
NeuroBayes (BF) & 68 & 9.52 &  56.3 \% \\
Decision Tree (AF) & & 7.73 &  137.0 \% \\
Decision Tree (BF) & & 7.53 &  151.5 \% \\
Random forest (AF) & & 10.72 &  23.1 \% \\
Random forest (BF) & & 9.22 &  66.5 \% \\
\end{tabular}
\caption{Comparisons of different ML techniques and algorithms using $\frac{S}{\sqrt{B}}$ as the figure of merit. The value of AxB for structure refers to A layers with B nodes per layer. For each classifier, a cut is placed on the network to maximize the value of the figure of merit. The value of $\frac{S}{\sqrt{B}}$ is shown for $100\ifb$. The last column is the relative amount of increase in data needed to match the (statistical-only) sensitivity of the best algorithm, shown in the top row. AF refers to the use of all features, EF refers to the use of extended features only, and BF refers to the use of basic features only. Only the best performing algorithms for each category are shown.}
\label{tab:compalgoSB}
\end{table*}

To understand the full shape of the output from these models, Figure~\ref{fig:NeuroBGDOutput} depicts signal and background distributions from NeuroBGD and XGBoost.
%To understand the full shape of neural network output, Figure~\ref{fig:NeuroBGDOutput} depicts the outputs of NeuroBGD model of both signal and background. Figure~\ref{fig:XGBoostOutput} shows output from the XGBoost model, which has similar performance. 
The output distributions from the NeuroBayes model are shown in Figure~\ref{fig:NeuroBayesOutput}. The structure seen is related to the $b$-tagging, with clear ``bumps'' arising from regions with different numbers of $b$-tags. The more complex algorithms shown in Figure~\ref{fig:NeuroBGDOutput} do not have these features. 

%\begin{figure*}[!htb]
%\begin{center}
%\trimmedgraphicsmall{XGBoost_AllFeatures_Output}
%\caption{Output distributions from XGBoost (FIXME - ADD TRAINING PARAMETERS), trained using all available features in 338K events. Only test data is shown in the plots.}
%\label{fig:XGBoostOutput}
%\end{center}
%\end{figure*}

\begin{figure*}[!htb]
\begin{center}
\trimmedgraphicsmall{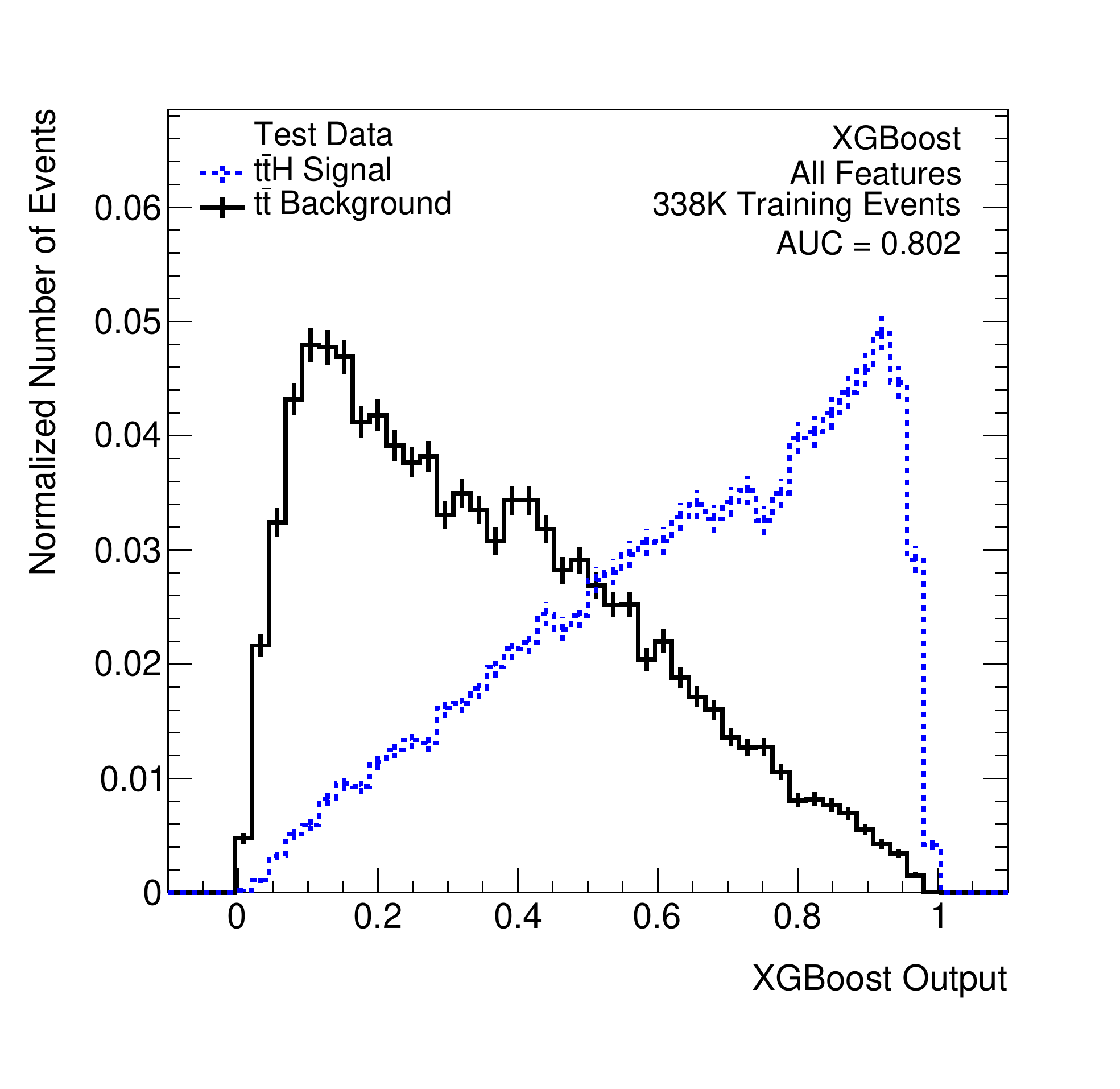}
\trimmedgraphicsmall{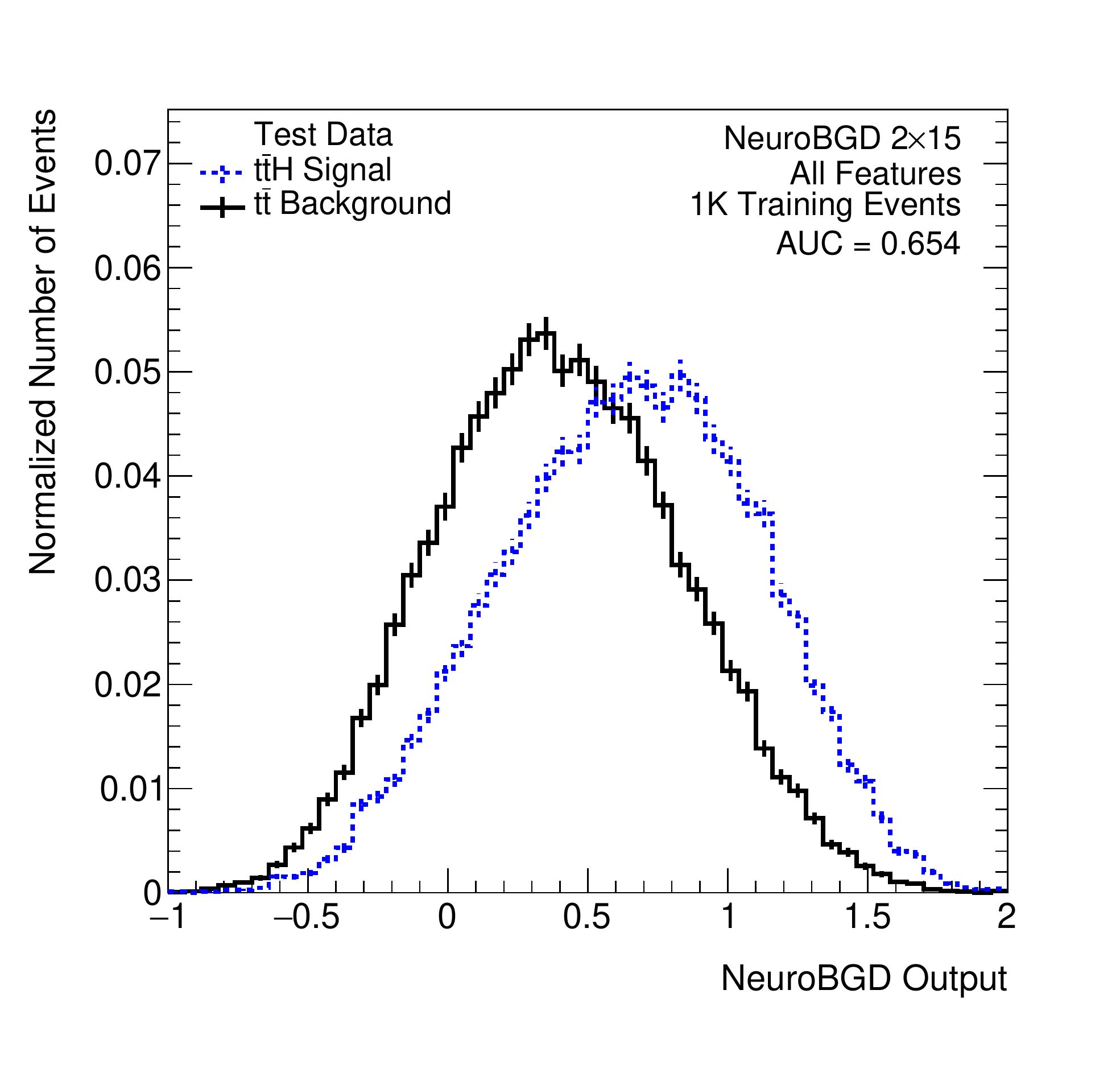}
\trimmedgraphicsmall{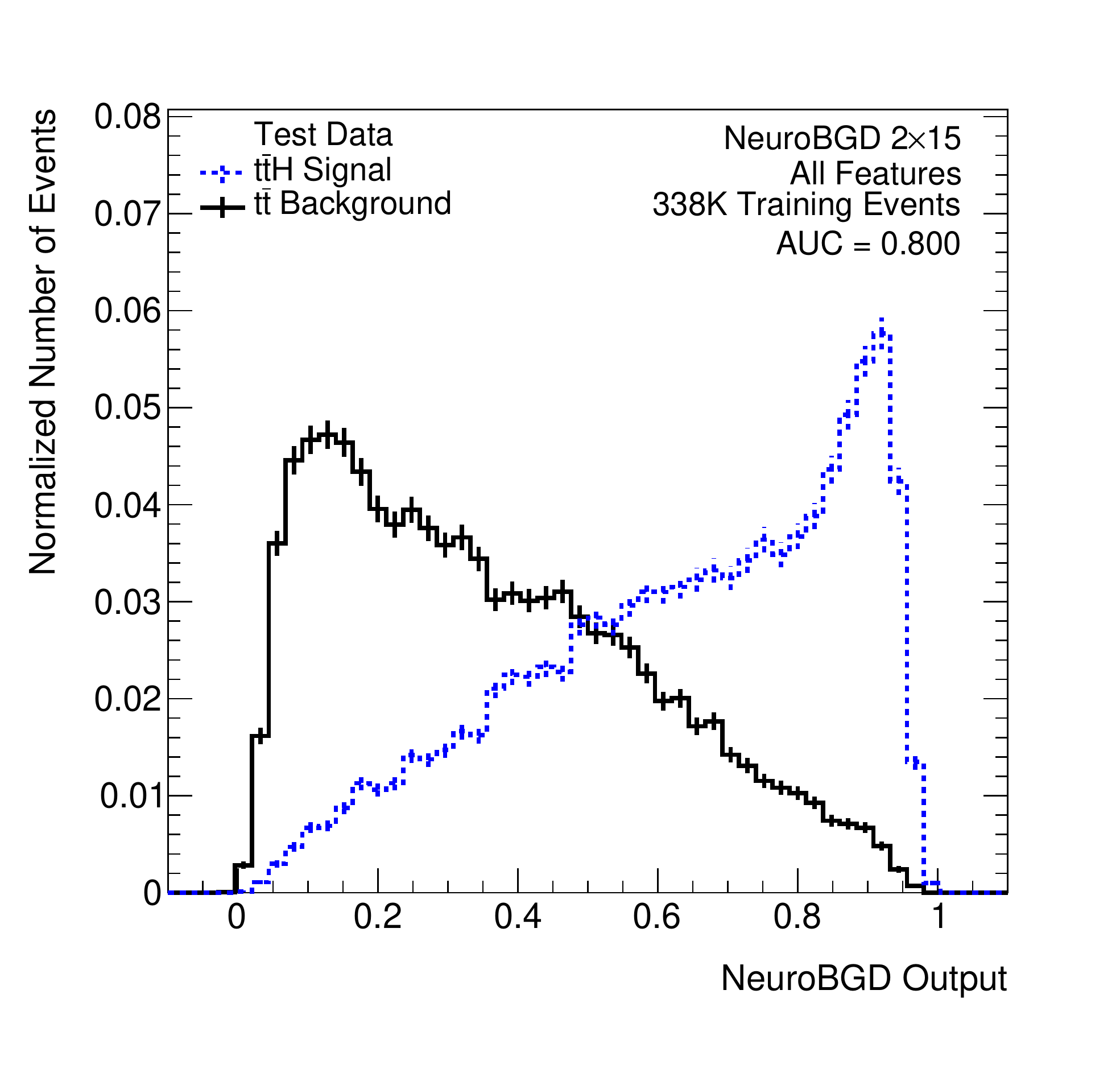}
\caption{Output distributions from XGBoost and from the NeuroBGD model with 2 layers and 15 nodes per layer. Left: XGBoost output trained using 338K events. Middle: NeuroBGD output trained with trained using 1K events. Right: NeuroBGD output trained using 338K events. Only test data is shown in these plots.}
\label{fig:NeuroBGDOutput}
\end{center}
\end{figure*}

\begin{figure*}[!htb]
\begin{center}
\trimmedgraphicsmall{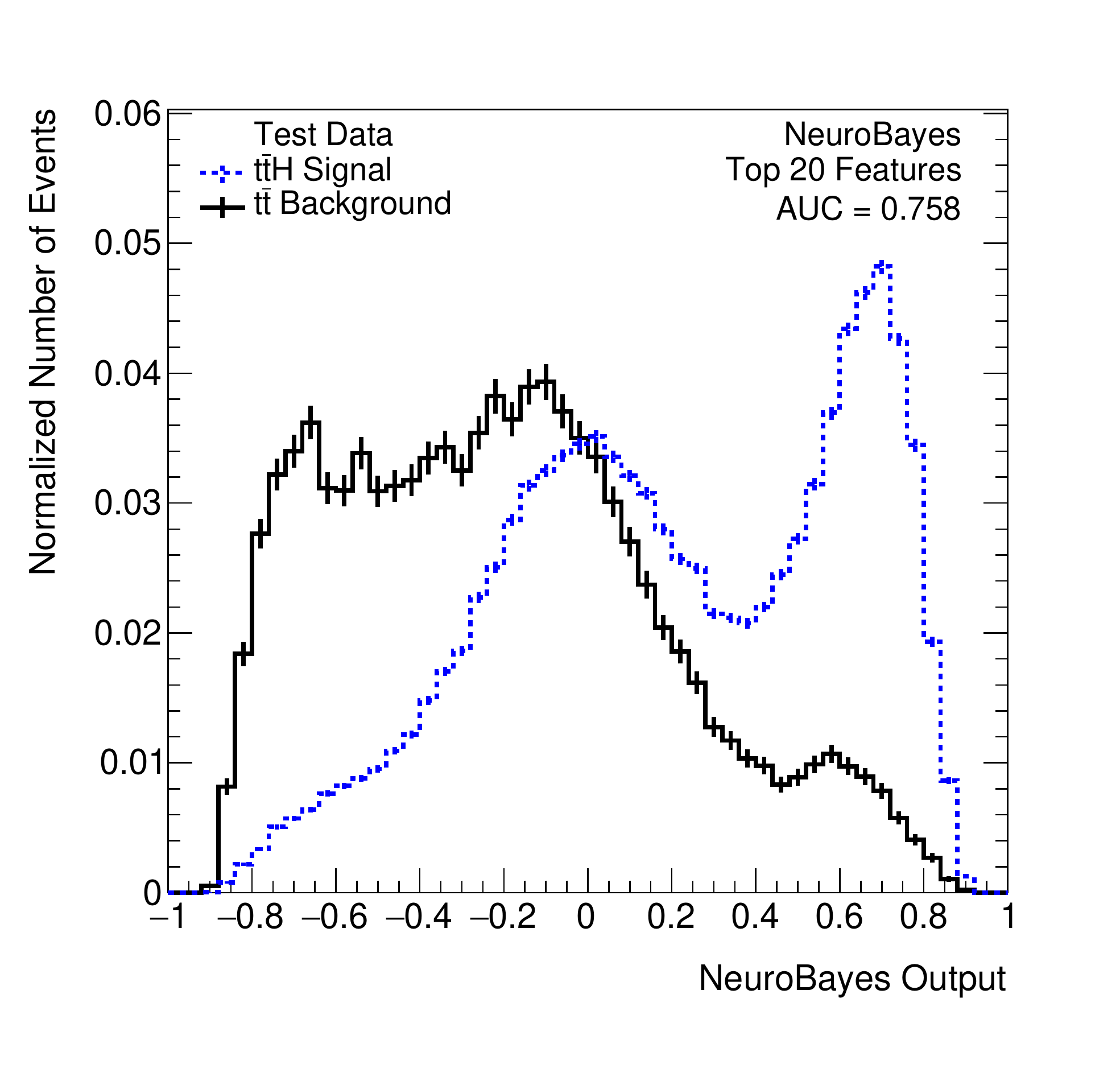}
\trimmedgraphicsmall{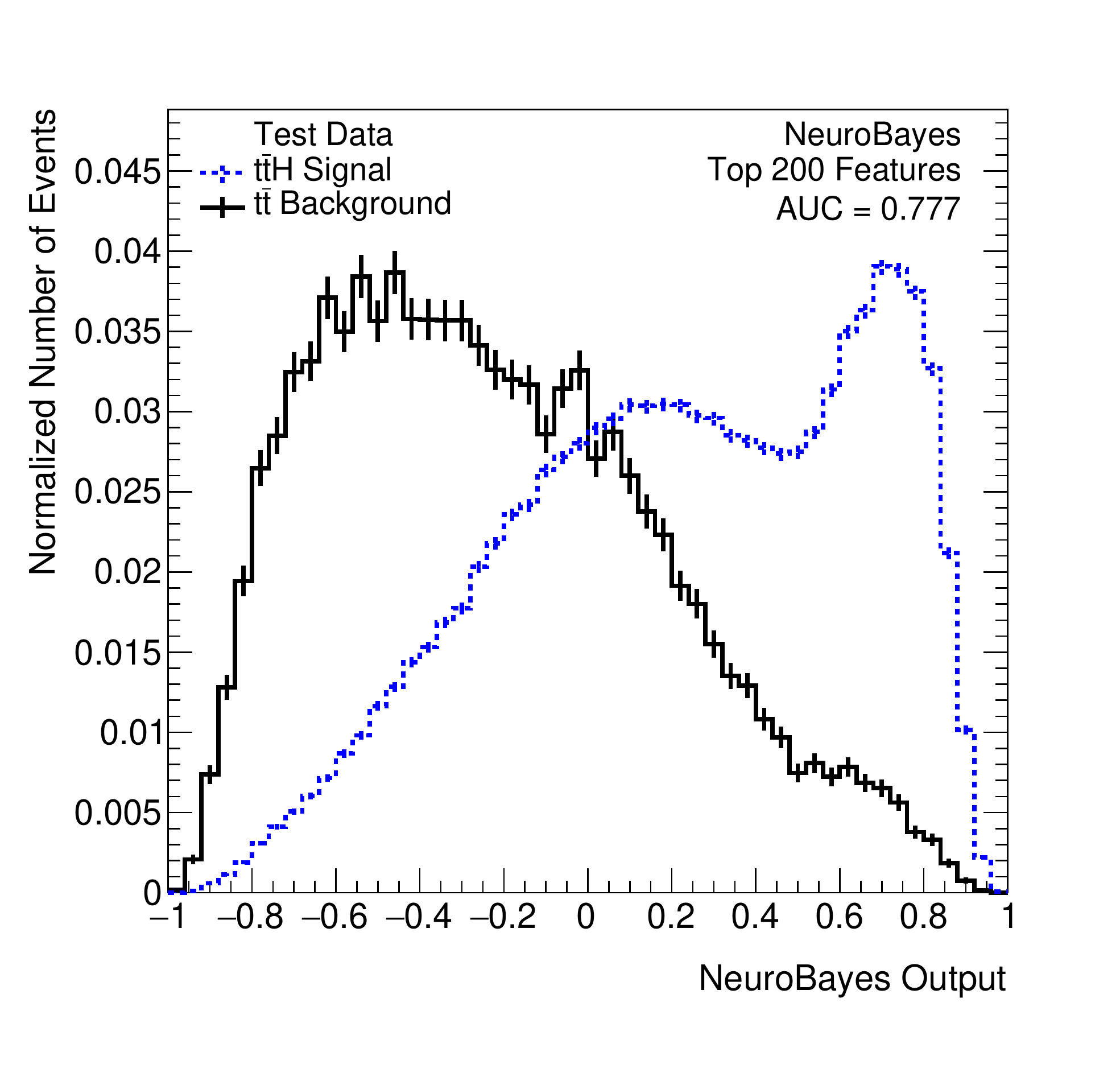}
\caption{NeuroBayes output distributions for the NeuroBayes MVA trained using the top 20 (left) and top 200 (right) discriminating variables}
\label{fig:NeuroBayesOutput}
\end{center}
\end{figure*}

\subsection{On features}  \label{sec:features}
Table~\ref{tab:fea} compares the results of using basic features, extended features and all features. For this comparative experiment, the NeuroBGD classifier is used with the data sets as described in Section~\ref{sec:algocomp}.
%As can be seen by an increase of 4-5\% AUC and large improvements in $\frac{S}{\sqrt{B}}$ on all tested neural structures, using the set of extended features is shown to be more effective than using the basic feature only.
Using the set of extended features is shown to be more effective than using basic features, as can be seen from the 4-5\% increases in AUC and the large improvements in $\frac{S}{\sqrt{B}}$ for all tested neural structures.
The complete set of features (basic and extended variables) further increase the AUC by 1\% and the $\frac{S}{\sqrt{B}}$ figure of merit as well.  Figure~\ref{fig:ROCCurves1} (right plot) shows the ROC curves of a NeuroBGD classifier when partial or all features are used for learning.  

The same trend is observed with the XGBoost classifier:  The highest performance of 80.2\% AUC is obtained with all features, while the basic feature only delivers 73.6\% AUC and the extended feature set delivers 79.6\% AUC, all using the full training set.  The results support the effectiveness of using or including the extended feature set.

One interesting observation is that when only basic features are used, NeuroBGD consistently delivers a better AUC than XGBoost.  However, when the extended features are added, XGBoost delivers comparable or higher AUC. The comparison is shown in Table~\ref{tab:XGvsNeu}.  The better performance of NeuroBGD when using only basic features can be related to its capability to extract correlations and information from these variables for classification. Once extended features are involved, the effect of NeuroBGD for feature extraction becomes less critical since the extended features are already the combination of basic features, based on physical principles.

\begin{table*}[!t]
\center
     \begin{tabular}{  c | c    c    c  }
     \hline
    \# of Hidden Layers x Nodes per Layer & { Basic Features} & {Extended Features} & { All Features }    \\ \hline
    2 x 15 & 74.4 &  79.7 & 80.0 \\ \hline
    2 x 300 & 74.3& 79.0 & 79.6 \\ \hline
   5 x 300 & 74.3 &  78.9 & 79.6 \\ \hline
   8 x 300&  74.3&  78.7 & 79.5 \\ \hline
   11 x 300 & 74.3 &  78.8 &  79.5 \\ \hline
\end{tabular}
\caption{Performance of \tth\  detection using different sets of features for the NeuroBGD model (AUC in \%) . }
\label{tab:fea}
\end{table*}

\begin{table*}[!t]
\center
     \begin{tabular}{  c | c    c |   c  c  | c  c }
     \hline
    training size & \multicolumn{2}{c|}{ Basic Features} & \multicolumn{2}{c|}{Extended Features}  &  \multicolumn{2}{c}{ All Features }   \\ \hline
      & { XGBoost} & {NeuroBGD} & {XGBoost} &{NeuroBGD} &  {XGBoost}  &  {NeuroBGD}   \\ \hline
   1k   & 63.8 & 66.5  & 70.4  & 65.7 & 70.8 & 66.5 \\ \hline
   3k  & 66.9 & 70.2 &  71.8 & 70.8 & 72.2 & 70.2 \\ \hline
   10k  & 70.3 & 71.6  & 74.1  &  75.2 & 74.3 & 71.6\\ \hline
   30K & 72.1 & 72.5  &  76.8 & 76.8 & 77.2 & 72.5 \\ \hline
   100K & 73.2 & 73.8 &  78.7 & 78.2  & 79.2 & 73.8 \\ \hline
   338K & 73.6 & 74.4 &   79.6 & 79.7 & 80.2 & 80.0 \\ \hline
\end{tabular}
\caption{Performance comparison of of XGBoost and NeuroBGD on different feature sets (AUC in \%) . }
\label{tab:XGvsNeu}
\end{table*}

\begin{figure*}[!htb]
\begin{center}
\trimmedgraphicsmall{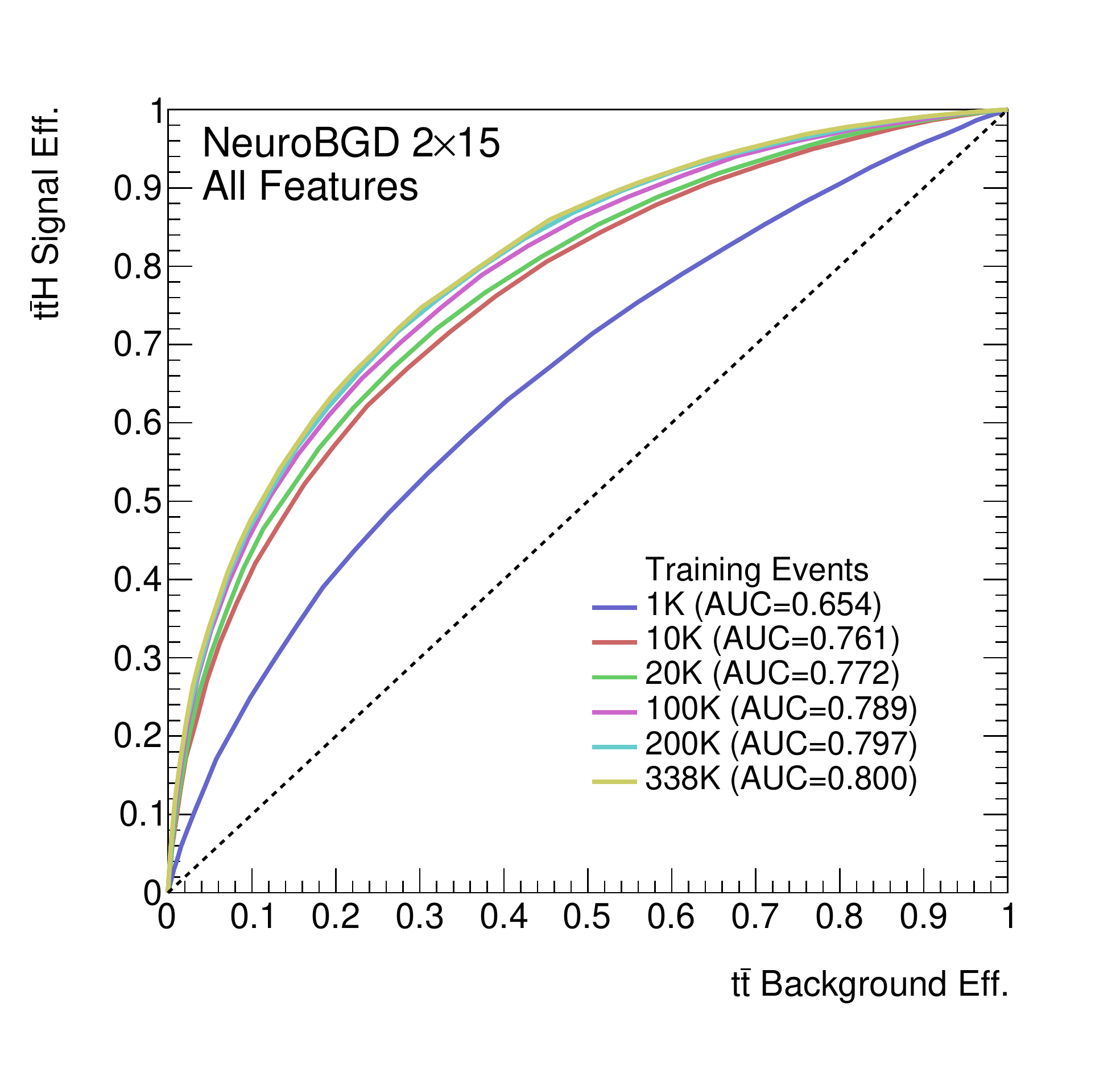}
\trimmedgraphicsmall{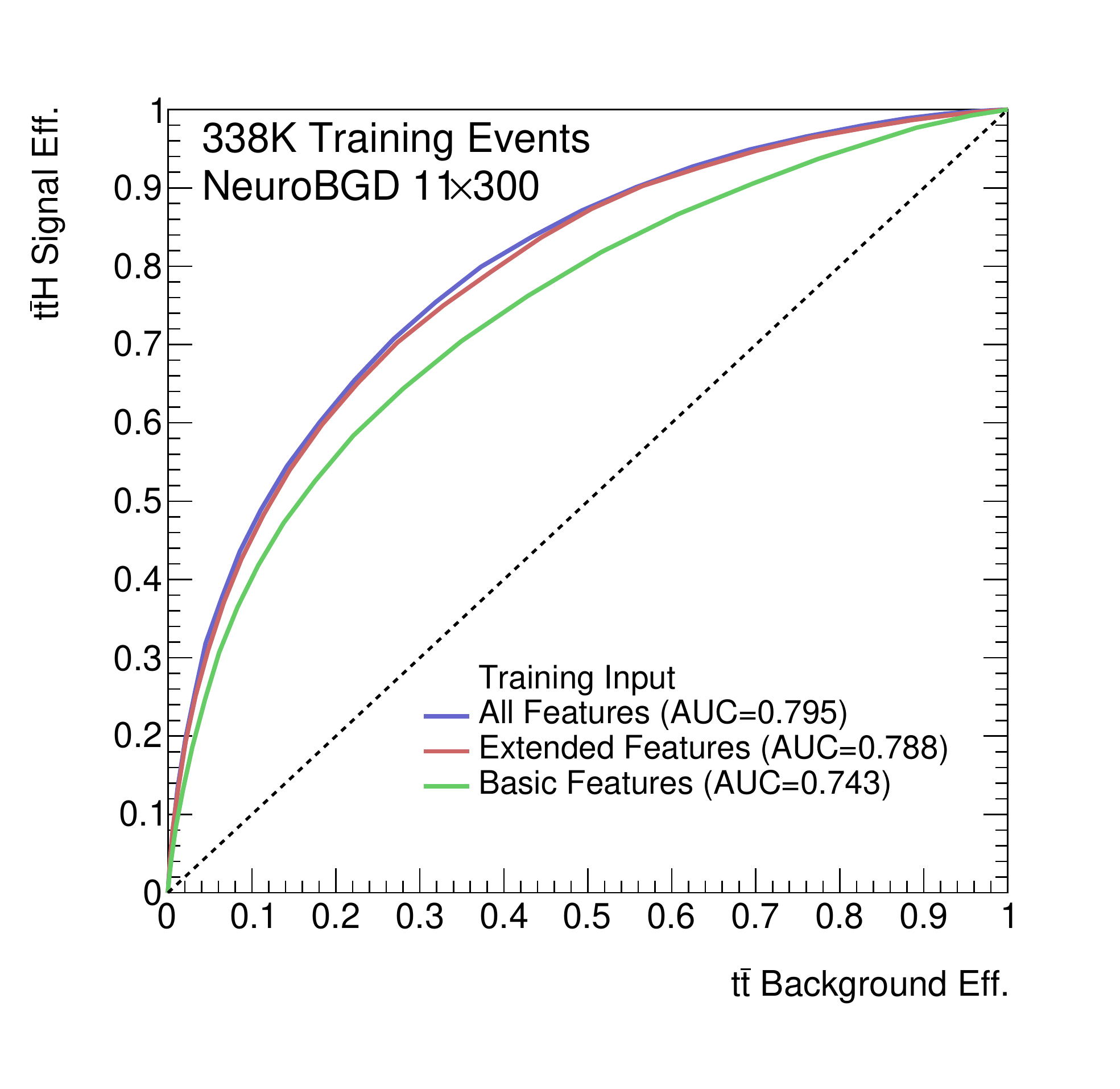}  
\caption{ROC curves for a variety of different NeuroBGD configurations. Left: results from a $2\times15$ network trained using all features with different training sample sizes. Right: results from a $11\times300$ network trained on the full training sample using different inputs.}
\label{fig:ROCCurves1}
\end{center}
\end{figure*}

To understand the effectiveness of features used as the inputs for the learning algorithms, we also calculate the AUC of each individual feature. The 597 features are then ranked using the individual AUC. As expected, $b$-tagging related variables are among the top ranked features, particularly the number of $b$-tagged jets passing a given threshold, as well as other variables such as the kinematics of $b$-tagged jets.

\subsection{On data size and training time}

It is important to understand the effect of the training set size for our problem, not only because data size often impacts learning, but also because of the difficult and lengthy process of simulating background events for this analysis. Figure \ref{fig:BarPlotsAF} shows how the performance changes as the size of the training set increases. The experiments were performed using the NeuroBGD classifier. Figure \ref{fig:ROCCurves1} (left plot) shows the ROC curves for the 2x15 network.

\begin{figure*}[!htb]
\begin{center}
\trimmedgraphicsmall{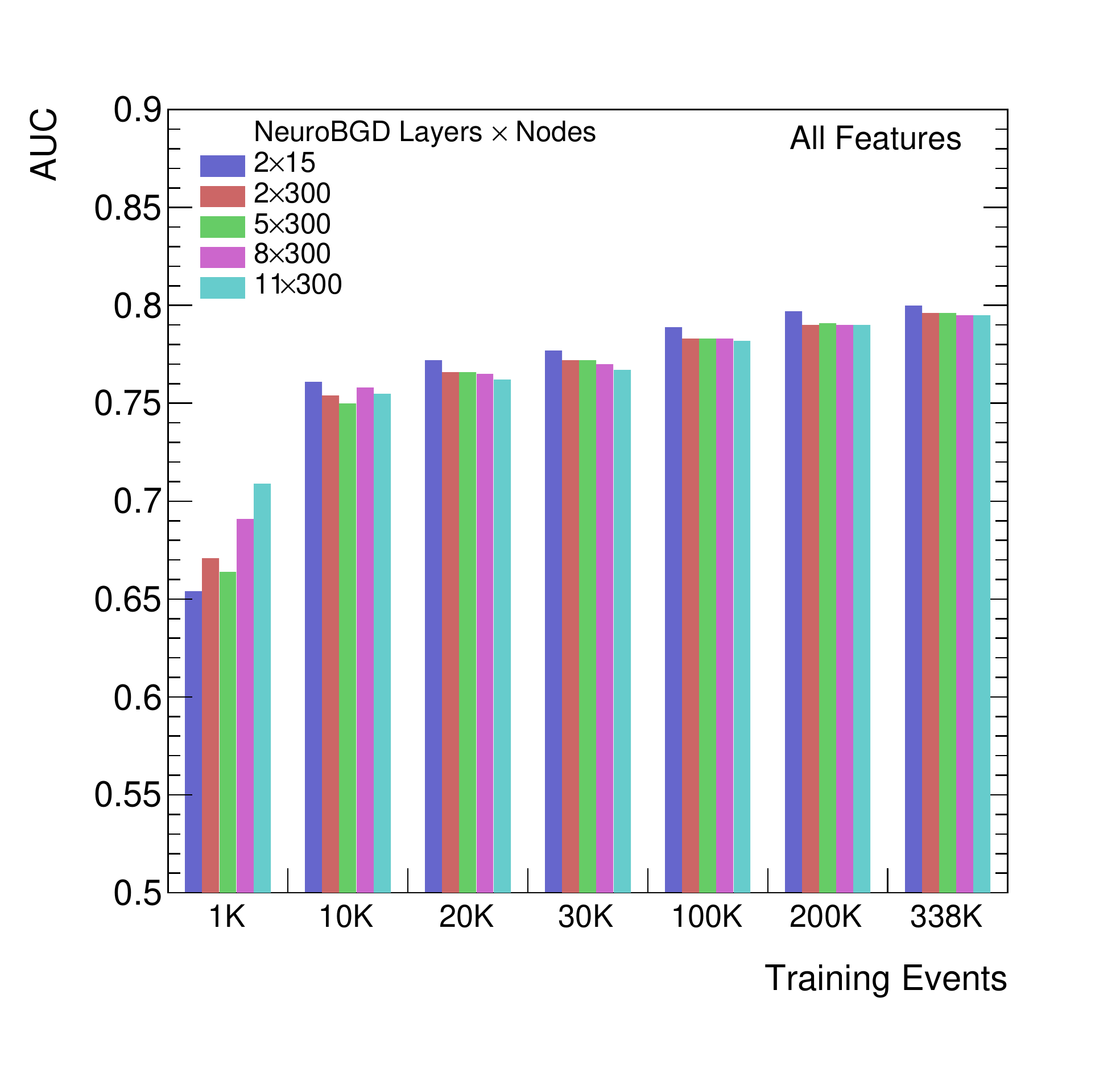}
\trimmedgraphicsmall{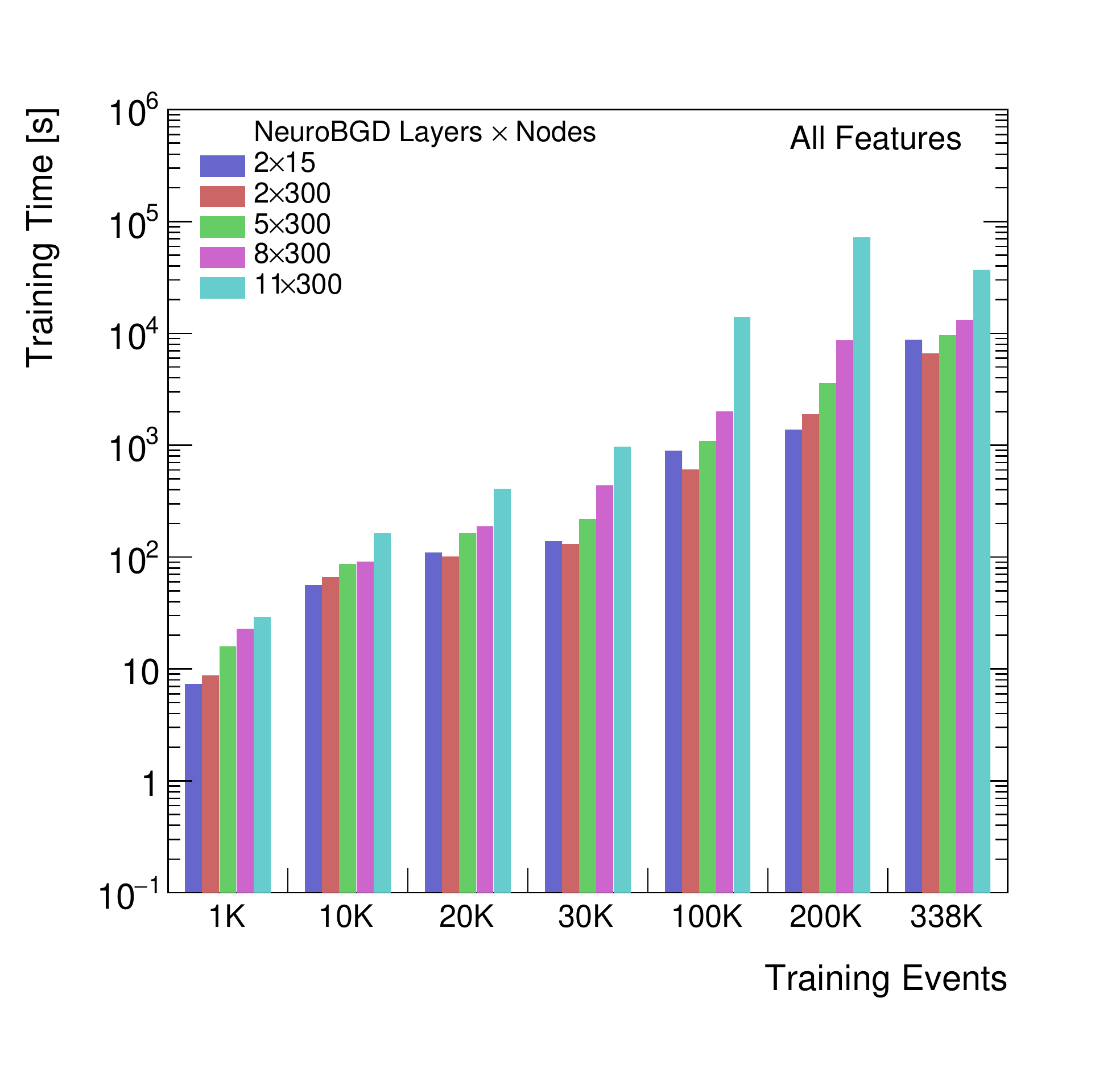}  
\caption{Summary of the AUC (left) and the total training time required (right) for all of the NeuroBGD networks trained on the full set of input variables. The vertical AUC axis in the left plot starts at 0.5 since this is the value that can be obtained when there is no separation between signal and background. The training times reported in the right plot vary for repeated experiments.}
\label{fig:BarPlotsAF}make
\end{center}
\end{figure*}

Not surprisingly, the AUC improves as the training set size increases.  For example, for the NeuroBGD with 2 layers and 15 nodes per layer, when the training set size increases from 1K to 338K, the AUC for detecting \tth\ increases from 65.5\% to 80.0\% (as is illustrated in Figure~\ref{fig:NeuroBGDOutput} and Figure~\ref{fig:BarPlotsAF}). The trend of AUC improvement plateaus as the training size continues toward the maximum value of 338K. Meanwhile, the time needed for training still grows at a significant rate as the training set size increases, from 17 seconds to more than 3 hours. Similar trends are observed for all tested structures of the neural classifier.  

For the XGBoost classifier,  when the training set size increases from 1K to 338K, the AUC for detecting \tth\ increases from 70.8\% to 80.2\%, with training time growing from 11 seconds to 324 seconds. XGBoost is significantly more efficient than NeuroBGD when the data size grows larger. When column subsampling is not used for XGBoost, the training time is about three times longer (about 15 minutes using 338K training events) but the AUC is not affected.

\subsection{On deep networks}

Recently in the ML field, there has been a strong interest in deep learning.  Deep neural networks, particularly deep convolutional neural networks (CNN) and variants have obtained many successes  \cite{AlexKNIPS2012}\cite{GoogleInception}.

To understand the impact of network depth for this problem, NeuroBGD of varying structures are tested. Results are shown in Figure~\ref{fig:ROCCurveStructure} and in Table \ref{tab:fea}. In this analysis, improvement in detection performance is not observed in deeper networks with more layers.  Networks of different number of layers deliver comparable AUCs, with the highest AUC obtained by a network of 2 layers and 15 nodes per layer (80.0\%).  In contrast, the networks of 8 or 11 layers and 300 nodes per layer deliver an AUC of 79.5\%.

\begin{figure}[!htb]
\begin{center}
\trimmedgraphic{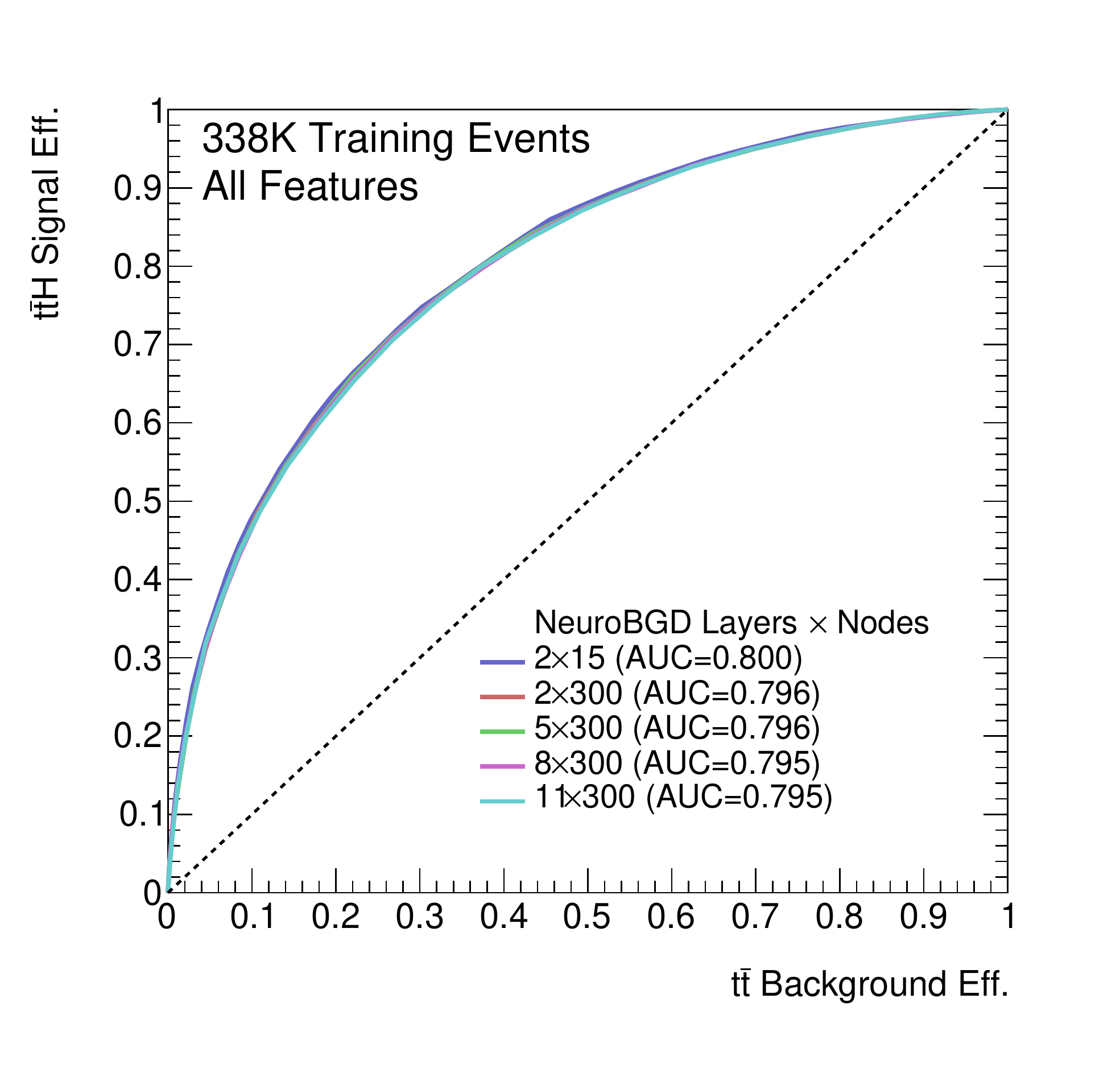}
\caption{ROC curves from NeuroBGD networks with different numbers of layers and nodes per layer. The variation in performance is small so the curves overlap.}
\label{fig:ROCCurveStructure}
\end{center}
\end{figure}

For XGboost, it is also found the the best AUC is obtained with tree depth of 1.  When deeper trees with depths of 2 to 6 are tested with the same parameter setup and number of iterations, lower AUCs between 74.9\% and 79.3\% are obtained and overfitting is observed.

The observation that deep model is not helpful for the learning problem seems surprising but is not entirely unexpected.  Among the applications in deep learning, the most successful domains have been with {\em unstructured data}, which refer to inputs with spatial or temporal correlations such as images and speech. For those problems, explicit feature extraction from the raw images is often not required, especially when models such as convolutional neural networks are employed. Such deep networks also demand a huge training set (often in millions) and a long training time. For domains and problems where much effort has put into carefully crafting features (referred to as feature engineering) and thus that have {\em structured data}, alternative learning methods can deliver comparable performance with reduced computational complexity in training, as has been found among Kaggle ML competitions \cite{chen2016xgboost}.  

Our empirical results suggest that advanced ensembles of trees and neurons, not necessarily deep, work effectively for the problem, or at least for the given features. With the reason still to be further explored, one possible explanation of such phenomenon is that the discriminating information for detecting \tth\ reside in the individual features, especially in the extended features that are inspired by physics and already contain the combination of pieces of information.  This explanation is supported by the fact that when only basic features are used, NeuroBGD works better than XGBoost across the board on all training data sizes. However, they become comparable once extended features are introduced. Meanwhile, there also exists a possibility that the advantage of a deeper model is yet to be revealed with larger training sets to offset overfitting.

\subsection{Future work}

The detection of \tth\ remains an important but difficult learning problem, though multiple avenues exist to improve the separation of signal and background in future studies.  One potential improvement, as-of-yet studied with deep networks, is the use of expert classification tools, where initial classifiers are used to make jet assignments to objects from the collision and thus to remove combinatorial issues. The information from these initial classifiers would then be used in a second ML algorithm. Such techniques were used in the recent ATLAS \tth\ search~\cite{ATLAS-CONF-2016-080}, and offered significant improvement over single classifier techniques.

\section{Conclusion}
The use of ML algorithms has already been shown to be important for observing rare particle physics processes such as \tth, though the use of more sophisticated and advanced techniques offer important gains. A clear example of the use of such tools is shown in Figure~\ref{fig:StackedMbb}, which indicates the power of a neural network to separate out signal and background. Before any ML requirements, the small signal is extremely difficult to see on top of a very large background; after placing a cut on a neural network discriminant, the signal becomes clearly visible by eye.

In the data sets studied here, both NeuroBGD and NeuroSGD outperform NeuroBayes along with the four other methods tested, both in terms of AUC, as well as in terms of ``equivalent data needed for discovery.'' Furthermore, it is shown that XGBoost has similar performance to NeuroBGD and NeuroSGD, but takes less training time compared to the NNs. The availability of large samples for training is found, as expected, to be important for network training. The benefit of deeper models, however, is not proved for our problem in either NeuroBGD or XGBoost, two of the top performing models. 

 The use of both extended and basic features in combination is found to give marginally better performance than either single feature set. When only basic features are used, the neural networks are found to perform better than XGBoost, likely because they are able to reproduce some of the extended features; once the extended features are included, this performance difference vanishes, indicating that the standard set of physics-inspired features captures most of the salient information that separates signal from background.

\begin{figure*}[!htb]
\begin{center}
\trimmedstack{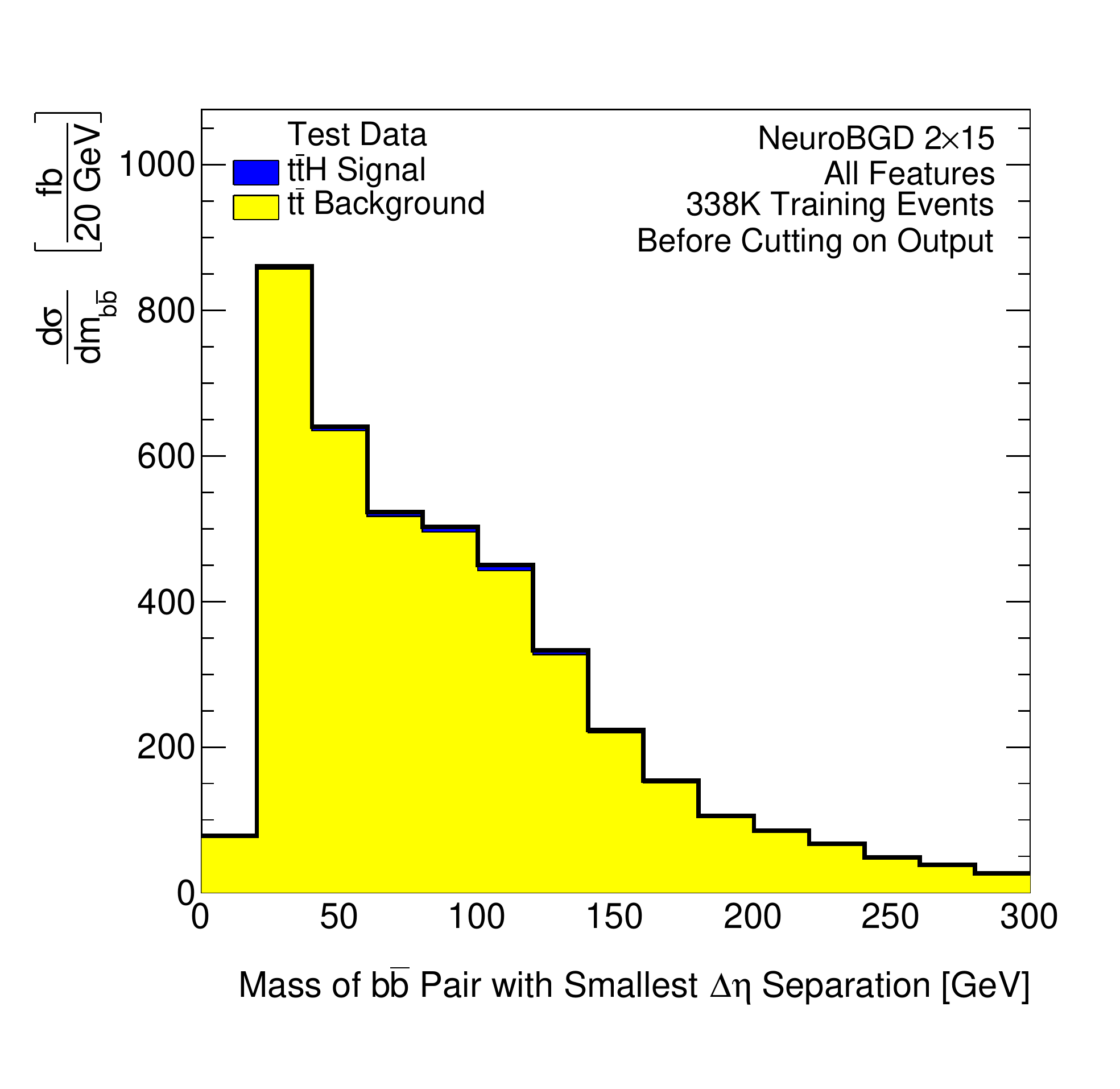}
\trimmedstack{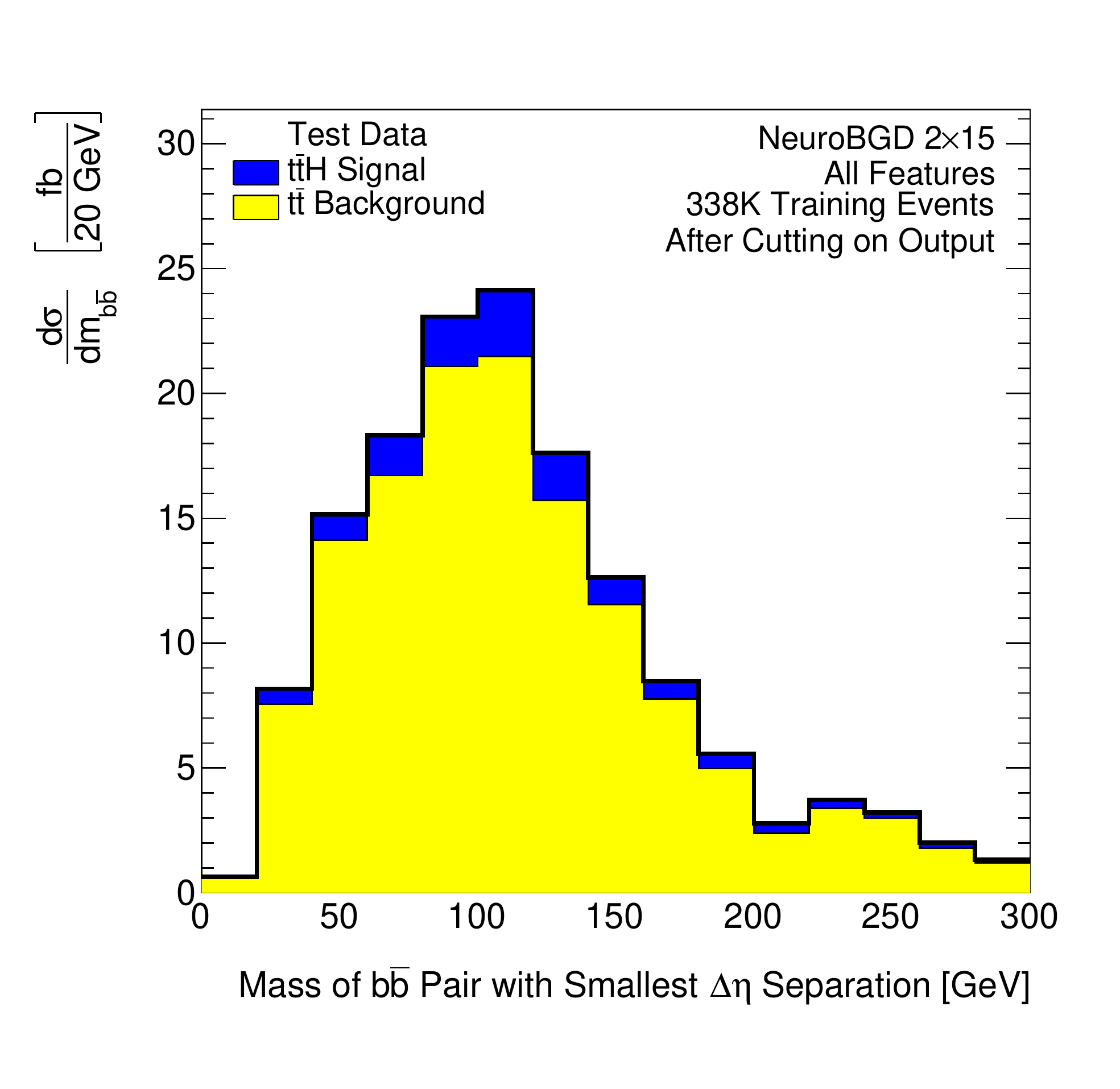}
\caption{Stacked signal and background plots of a candidate reconstructed Higgs mass variable, normalized to their respective cross-sections. Left: Signal and background after the loose event selection described in section~\ref{lab:eventSel} is applied, but before any ML requirement. The signal contribution is difficult to see due to the large background. Right: The same simulated data after placing a cut on the NeuroBGD output (using the $2\times15$ structure shown on the right in Fig~\ref{fig:NeuroBGDOutput}) to maximize $S/\sqrt{B}$. After cutting the signal becomes clearly visible. The background also peaks near the Higgs mass of 125~GeV due to making a requirement that events look signal-like.}
\label{fig:StackedMbb}
\end{center}
\end{figure*}

\section{Acknowledgments and support}
JA is supported by National Science Foundation Grant 1505989. Argonne National Laboratory's work was supported by the U.S. Department of Energy, Office of Science under contract DE-AC02-06CH11357. The authors are grateful to Daniel Whiteson and Taylor Childers for their helpful suggestions and feedback, and thank Dr. Michael Papka of DDI lab at NIU for providing the hardware resources for this research.

\bibliography{paper}

\end{document}